\documentclass{aa}
\usepackage[utf8]{inputenc}
\usepackage[varg]{txfonts}
\usepackage[breaklinks, colorlinks, citecolor=blue, linkcolor=blue]{hyperref}
\usepackage[usenames, dvipsnames]{color}
\usepackage{amsmath}
\usepackage{graphicx}
\usepackage[english]{babel}
\usepackage{siunitx}
\usepackage{float}
\usepackage{url}
\usepackage{longtable}
\usepackage{fancyhdr}
\usepackage{footmisc}
\usepackage{natbib}
\usepackage[normalem]{ulem}
\usepackage{caption}
\usepackage{multirow}
\usepackage{color}
\usepackage{array} 
\usepackage{ulem}
\usepackage[flushleft]{threeparttable}

\makeatletter
\def\instrefs#1{{\def\scsep{\def\scsep{,}}\@for\w:=#1\do{\scsep\ref{inst:\w}}}}
\renewcommand{\inst}[1]{\unskip$^{\instrefs{#1}}$}

\renewcommand*\aa@pageof{, page \thepage{} of \pageref*{LastPage}} 

\makeatother

\title{CARMENES detection of the \ion{Ca}{ii} infrared triplet \\ and possible evidence of \ion{He}{i} in the atmosphere of WASP-76b}

\titlerunning{CARMENES detection of \ion{Ca}{ii} IRT and possible evidence of \ion{He}{i} in the atmosphere of WASP-76b}

\author{N.~Casasayas-Barris\inst{lo}
        \and
        J.~Orell-Miquel\inst{iac,ull}
        \and
        M.~Stangret\inst{iac,ull}
        \and
        L.~Nortmann \inst{gott}
        \and
        F.~Yan\inst{gott}
        \and
        M.~Oshagh\inst{iac,ull}
        \and
        E.~Palle\inst{iac,ull}
        \and
        J.~Sanz-Forcada\inst{cab}
        \and
        M.~L\'opez-Puertas\inst{iaa}
        \and
        E.~Nagel\inst{ham,tls}
        \and
        R.~Luque\inst{iaa,iac,ull}
        \and
        G.~Morello\inst{iac,ull}
        \and 
        I.\,A.\,G.~Snellen \inst{lo}
        \and
        M.~Zechmeister\inst{gott}
        \and
        A.~Quirrenbach\inst{lsw}
        \and
        J.\,A.~Caballero\inst{cab}
        \and
        I.~Ribas\inst{ice,ieec}
        \and
        A.~Reiners\inst{gott}
        \and
        P.\,J.~Amado\inst{iaa}
        \and
        G.~Bergond\inst{caha}
        \and
        S.~Czesla\inst{ham}
        \and
        Th.~Henning\inst{mpia}
        \and
        S.~Khalafinejad\inst{lsw}
        \and
        K.~Molaverdikhani\inst{lsw,mpia,lmu}
        \and
        D.~Montes\inst{ucm}
        \and
        M.~Perger\inst{ice,ieec}
        \and
        A.~S\'anchez-L\'opez\inst{lo}
        \and
        E.~Sedaghati\inst{eso,iaa}
        }

\institute{
\label{inst:lo}Leiden Observatory, Leiden University, Postbus 9513, 2300 RA Leiden, The Netherlands; \email{barris@strw.leidenuniv.nl}
\and 
\label{inst:iac}Instituto de Astrof\'isica de Canarias (IAC), 38205 La Laguna, Tenerife, Spain
\and 
\label{inst:ull}Departamento de Astrof\'isica, Universidad de La Laguna (ULL), 38206, La Laguna, Tenerife, Spain
\and 
\label{inst:gott}Institut f\"ur Astrophysik, Georg-August-Universit\"at, Friedrich-Hund-Platz 1, 37077 G\"ottingen, Germany
\and
\label{inst:cab}Centro de Astrobiolog\'ia (CSIC-INTA), ESAC, Camino bajo del castillo s/n, 28692 Villanueva de la Ca\~{n}ada, Madrid, Spain
\and
\label{inst:iaa}Instituto de Astrof\'isica de Andaluc\'ia (IAA-CSIC), Glorieta de la Astronom\'ia s/n, 18008 Granada, Spain
\and
\label{inst:ham}Hamburger Sternwarte, Universit\"at Hamburg, Gojenbergsweg 112, 21029 Hamburg, Germany
\and
\label{inst:tls}Th\"uringer Landessternwarte Tautenburg, Sternwarte 5, 07778 Tautenburg, Germany
\and
\label{inst:lsw}Landessternwarte, Zentrum f\"ur Astronomie der Universit\"at Heidelberg, K\"onigstuhl 12, 69117 Heidelberg, Germany
\and
\label{inst:ice}Institut de Ci\`encies de l'Espai (CSIC-IEEC), Campus UAB, c/ de Can Magrans s/n, 08193 Bellaterra, Barcelona, Spain
\and 
\label{inst:ieec}Institut d'Estudis Espacials de Catalunya (IEEC), 08034 Barcelona, Spain
\and
\label{inst:caha}Observatorio de Calar Alto, Sierra de los Filabres, 04550 G\'ergal, Almer\'ia, Spain
\and
\label{inst:mpia}Max-Planck-Institut f\"ur Astronomie, K\"onigstuhl 17, 69117 Heidelberg, Germany
\and
\label{inst:lmu}Universit\"ats-Sternwarte, Ludwig-Maximilians-Universit\"at M\"unchen, Scheinerstrasse 1, 81679 M\"unchen, Germany
\and 
\label{inst:ucm}Departamento de F\'isica de la Tierra y Astrof\'isica and IPARCOS-UCM (Unidad de F\'isica de Part\'iculas y del Cosmos de la UCM), Facultad de Ciencias F\'isicas, Universidad Complutense de Madrid, 28040 Madrid, Spain
\and
\label{inst:eso}European Southern Observatory, Alonso de C\'ordova 3107, Santiago, Chile
           }

\date{Received 29 June 2021 / Accepted 26 August 2021}

\abstract{
Ultra-hot Jupiters are highly irradiated gas giants with equilibrium temperatures typically higher than 2000\,K. Atmospheric studies of these planets have shown that their transmission spectra are rich in metal lines, with some of these metals being ionised due to the extreme temperatures. Here, we use two transit observations of WASP-76b obtained with the CARMENES spectrograph to study the atmosphere of this planet using high-resolution transmission spectroscopy. Taking advantage of the two channels and the coverage of the red and near-infrared wavelength ranges by CARMENES, we focus our analysis on the study of the \ion{Ca}{ii} infrared triplet (IRT) at 8500\,{\AA} and the \ion{He}{i} triplet at 10830\,{\AA}. We present the discovery of the \ion{Ca}{ii} IRT at 7$\sigma$ in the atmosphere of WASP-76b using the cross-correlation technique, which is consistent with previous detections of the \ion{Ca}{ii} H\&K lines in the same planet, and with the atmospheric studies of other ultra-hot Jupiters reported to date. The low mass density of the planet, and our calculations of the XUV (X-ray and EUV) irradiation received by the exoplanet, show that this planet is a potential candidate to have a \ion{He}{i} evaporating envelope and, therefore, we performed further investigations focussed on this aspect.
The transmission spectrum around the \ion{He}{i} triplet shows a broad and red-shifted absorption signal in both transit observations. However, due to the strong telluric contamination around the \ion{He}{i} lines and the relatively low signal-to-noise ratio of the observations, we are not able to unambiguously conclude if the absorption is due to the presence of helium in the atmosphere of WASP-76b, and we consider the result to be only an upper limit. Finally, we revisit the transmission spectrum around other lines such as \ion{Na}{i}, \ion{Li}{i}, H$\alpha$, and \ion{K}{i}. The upper limits reported here for these lines are consistent with previous studies.}


\begin{document}

\maketitle

\section{Introduction}

Ultra-hot Jupiters are gas giants under extreme irradiation conditions. Due to their close-in orbits, these planets are strongly irradiated by their host stars and, consequently, they show extreme equilibrium temperatures, typically higher than 2000\,K \citep{Parmentier2018,Arcangeli2018}. At these temperatures, molecules become thermally dissociated and the atoms are ionised \citep{Helling2019}. The irradiation received in the upper atmosphere can produce significant atmospheric mass loss, which has a direct impact on the evolution of the atmosphere \citep{Fossati2018}, and it can additionally cause thermal inversions \citep{Lothringer2018}. The presence of inversion layers has been suggested to be due to the absorption of irradiation at short wavelengths by atomic metals, metal hydrides and oxides \citep{Pino2020}, the lack of coolants, such as water \citep{Parmentier2018,Molliere2015}, or the presence of clouds 
\citep[e.g.][]{Molaverdi2020}.
The permanent nightside of ultra-hot Jupiters, in contrast, could show relatively lower temperatures than the dayside and, therefore, a different atmospheric chemistry \citep{Arcangeli2018,Bell2018}. 

Several atoms and ions have been detected in the atmospheres of ultra-hot Jupiters, including Na, Fe, Fe$^+$, Mg, and Ca$^+$, among others (e.g. \citealt{Hoeijmakers2018,Stangret2020,Casasayas2019,Borsa2020}). Moreover, \citet{Keidberg2018} reported a lack of water in the day-side atmosphere of WASP-12b, \citet{YanKELT9} claimed strong H$\alpha$ absorption from the atmosphere of KELT-9b, indicating an escaping hydrogen atmosphere, and \citet{Ehrenreich2020} found the first observational indication of chemical differences between the day and night sides of the atmosphere of an ultra-hot Jupiter. Recently, emission of \ion{Fe}{i} has been detected in the dayside of three ultra-hot Jupiters \citep{Yan2020,Pino2020,Nugroho2020FeIe,Cont2021}, indicating the presence of thermal inversion in their atmospheres. 

Here, we analyse two transit observations of WASP-76b obtained with the CARMENES instrument (Calar Alto high-Resolution search for M dwarfs with Exoearths with Near-infrared and optical Échelle Spectrographs; \citealt{CARMENES,CARMENES20}), to explore the atmosphere of this planet. This study takes advantage of the coverage of CARMENES in the red-visual and near-infrared, and it is focussed on the analysis of the \ion{Ca}{ii} infrared triplet (IRT) around $8500$\,{\AA} and the \ion{He}{i} triplet at $10830$\,{\AA}. We additionally examine the \ion{Na}{i}, H$\alpha$, \ion{Li}{i}, and \ion{K}{i} lines for comparison with previous findings.

This paper is organised as follows. In Section~\ref{sec:wasp76} we introduce the WASP-76 system, in Section~\ref{sec:observations} we describe the observations, in Section~\ref{sec:analysis} we detail the analysis performed to extract the transmission spectrum, the results around the different lines are shown in Section~\ref{sec:results}, and the summary of the results and conclusions are presented in Section~\ref{sec:conclusions}.

\section{The WASP-76 system} \label{sec:wasp76}

WASP-76 (\object{BD+01~316}) is a system at a distance of about 195\,pc, composed of two stars and a planet that orbits the primary star (\citealt{Bohn2020}; hereafter we refer to the planet as WASP-76b, its host star as WASP-76, and the stellar companion as WASP-76B). WASP-76 is a bright ($V \approx9.5$\,mag, $K \approx 8.2$\,mag) F7-type star. The stellar companion, WASP-76B, has an effective temperature of $4800$\,K, which points to a late G- or an early K-type dwarf. It is $\Delta K = 2.30$\,mag, fainter than the primary, and it is separated from it by $0.436\pm0.003$\,arcsec \citep{Southworth2020}. 

\begin{figure}[]
\centering
\includegraphics[width=0.5\textwidth]{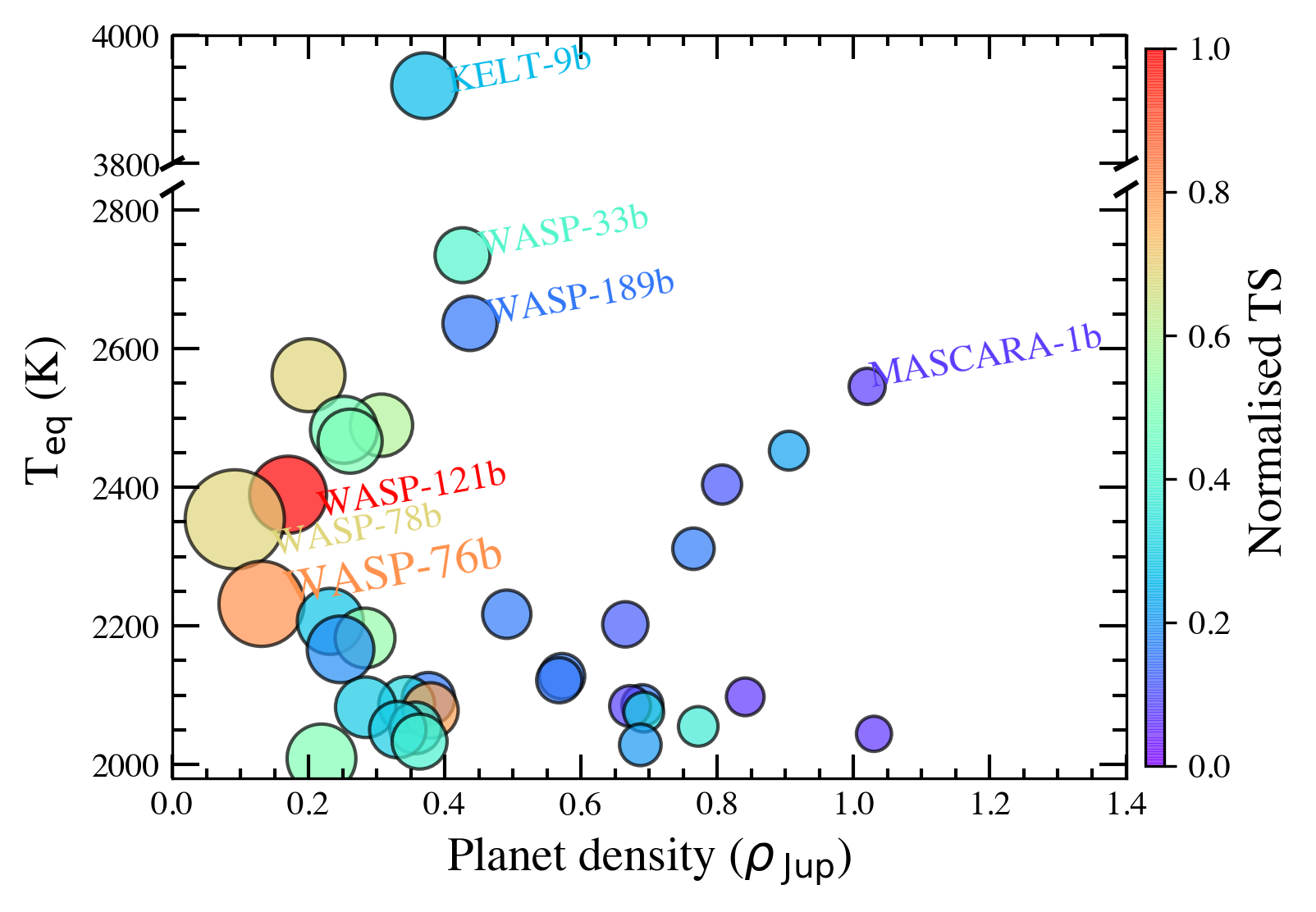}
\caption{Equilibrium temperature of known ultra-hot Jupiters ($T_{\rm eq}>2000$\,K) with measured mass calculated assuming zero albedo, represented as a function of the planet density. The colour bar shows the ratio of the exoplanet atmosphere annulus surface to the stellar disc area (TS) normalised to this particular sample, so the planet with the largest transmission has a value equal to unity. We assumed a mean molecular weight $\mu=2.3$ when computing the planet scale height, $H$. The symbol size of the markers is proportional to $H$ of each planet. All these values were calculated with the information extracted from the TEPCat catalogue \citep{TEPCat}.}
\label{fig:UHJ_context}
\end{figure}

With a radius of $1.85\,R_{\rm Jup}$ and a mass of $0.89\,M_{\rm Jup}$ \citep{Ehrenreich2020}, WASP-76b is one of the most inflated ultra-hot Jupiters (see Fig.~\ref{fig:UHJ_context}). 
It has an equilibrium temperature $T_{\rm eq} \approx 2160$\,K and orbits its host star every 1.81\,d \citep{West2016}. 
Its atmospheric scale height, $H$, assuming a mean molecular weight $\mu=2.3$, is $\sim1250$\,km, which, together with all these properties, make WASP-76b an excellent planet for atmospheric studies using transmission spectroscopy. 

The atmosphere of WASP-76b has been studied with high-resolution spectroscopy at different facilities. For example, \citet{Seidel2019} and \citet{Zak2019} detected \ion{Na}{i} in the atmosphere of WASP-76b using the HARPS spectrograph. \citet{Zak2019} additionally reported a non-detection of H$\alpha$ and H$\beta$. Using ESPRESSO, \citet{Ehrenreich2020} found \ion{Fe}{i} in the atmosphere of this planet and, in particular, reported an asymmetric atmospheric signal that can be explained by the combination of planetary
rotation and wind blowing from the dayside. In addition, this \ion{Fe}{i} signal was observed to arise mainly from the dayside, while disappearing close to the morning terminator, as Fe condenses in the night-side atmosphere. More recently, \citet{Tabernero2020} reported the detection of several species, including Na, Fe, Ca$^{+}$, Li, Mg, and Mn, also with ESPRESSO.

\section{Observations} \label{sec:observations}

\subsection{CARMENES observations} \label{sec:obs_carm}

Two transits of WASP-76b were observed with CARMENES, on the nights of 3 October and 29 October 2019. CARMENES is installed at the 3.5\,m telescope at the Calar Alto Observatory, and consists of two high-resolution spectrographs: the visible (VIS) channel covers the wavelength range from 0.52 to 0.96\,${\mu}$m and has a resolving power of $\mathcal{R} = 94\,600$, and the near-infrared channel (NIR) covers the range from 0.96 to 1.71\,${\rm \mu}$m with $\mathcal{R} = 80\,400$. Both transits were observed following the same strategy, this is, taking continuous exposures of WASP-76, starting before the transit and ending after the transit. Fibre A was used to observe the target and fibre B to monitor the sky. The size of CARMENES fibres on the sky is $1.5$\,arcsec, which means that both WASP-76 and its stellar companion are included in the fibre (see discussion in Section~\ref{sec:companion}). 
The individual exposure time were 492\,s in the VIS channel,and 498\,s in the NIR channel, obtaining the same number of exposures in both channels and a total of 44 and 40 exposures on the first and second night, respectively. During the transit, 25 and 24 exposures were obtained for the VIS and NIR channels, respectively. 

\begin{figure}[]
\centering
\includegraphics[width=0.5\textwidth]{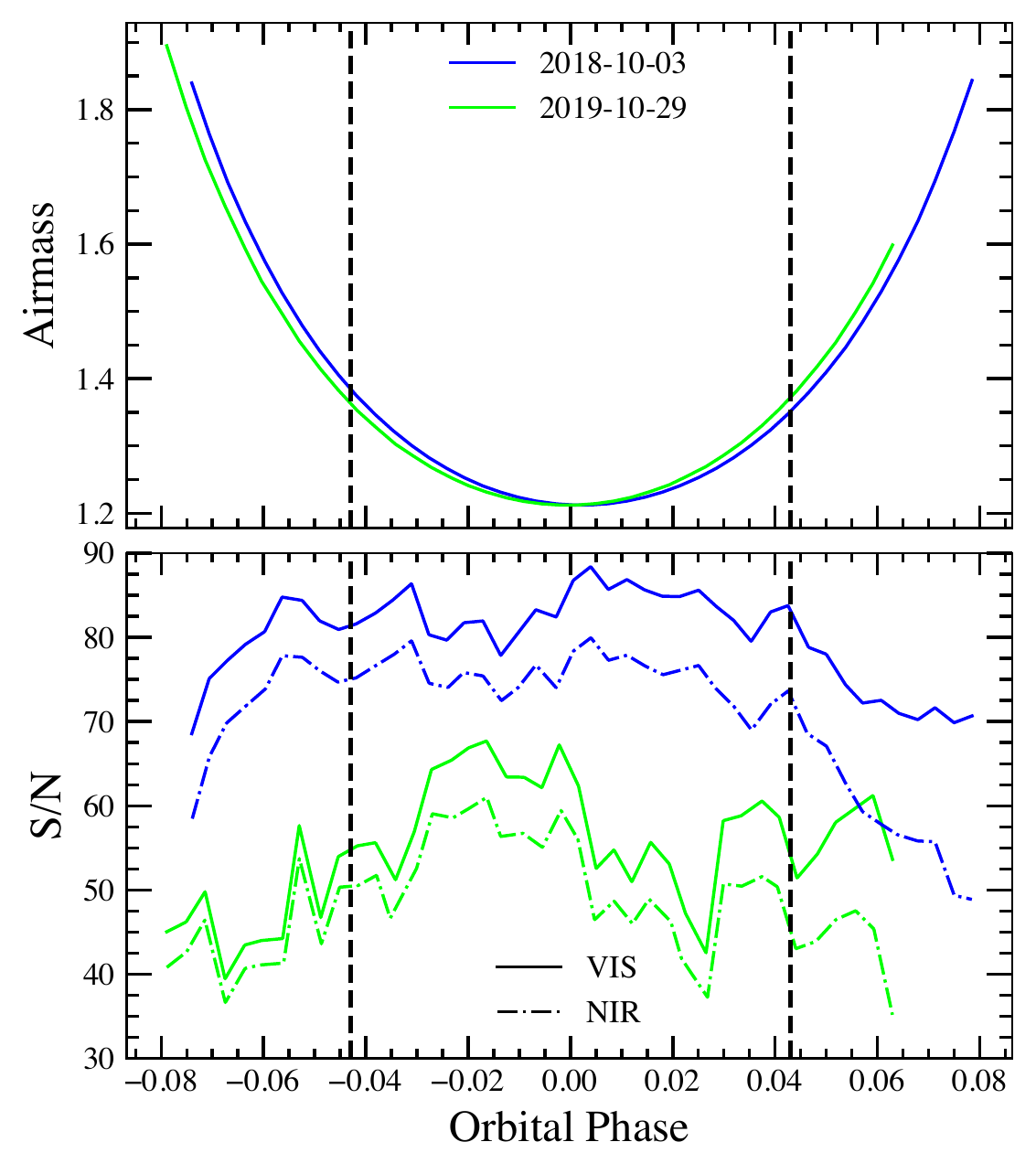}
\caption{Evolution of the airmass ({\it top panel}) and S/N ({\it bottom panel}) of the transit observations on Night 1 (blue) and Night 2 (green). In the bottom panel, the solid lines show the S/N in the VIS channel and the dot-dashed lines indicate the S/N in the NIR channel. The vertical-dashed lines indicate the first and last contacts of the transit.}
\label{fig:S/N}
\end{figure}

The observations are processed with the CARMENES pipeline {\tt caracal} (CARMENES Reduction And Calibration; \citealt{CARACAL}). This reduction considers bias removal, flat-relative optimal extraction \citep{Zechmeister2014}, cosmic ray correction, and wavelength calibration described by \citet{carmcalibration}. The reduced spectra are presented in the terrestrial rest frame and the wavelengths are given in vacuum. The mean signal-to-noise (S/N) of the two nights is 80 and 55 around $8500$\,{\AA} in the VIS channel, and 71 and 48 around $10900$\,{\AA} in the NIR channel, respectively. The S/N evolution during the nights is shown in Fig.~\ref{fig:S/N}, and the observations are summarised in Table~\ref{Tab:Obs}. 

\begin{table*}[]
\centering
\caption{Observing log of WASP-76b transit observations obtained with CARMENES.}
\begin{tabular}{ccccccccc}
\hline
\hline
\noalign{\smallskip}
Night & Date of   & Start & End &  Airmass & $N_\mathrm{obs}$ & Channel & $t_\mathrm{exp}$  & S/N$^b$ \\
         & observation & [UT] & [UT] & range$^a$ & & & [s] &   \\
\noalign{\smallskip}
\hline
\noalign{\smallskip}
1 & 2018-10-03 & 21:48 & 04:26 & 1.84--1.21--1.84 & 44 & VIS & 492 & 69--88 \\
 & & & & & & NIR & 498 & 49--80 \\
2 & 2019-10-29 & 20:01 &  02:11 & 1.89--1.21--1.60 & 40 & VIS & 492 & 39--68 \\ 
 & & & & & & NIR & 498 & 35--61 \\
\noalign{\smallskip}
\hline
\end{tabular}\\
\tablefoot{\tablefoottext{a}{Airmass range during the observation. The first and last values correspond to the airmass at the first and last exposures of the night. The second value is the minimum airmass near the culmination. } \tablefoottext{b}{Minimum and maximum S/N for each night, calculated around $8500$\,{\AA} for the VIS channel, and around $10900$\,{\AA} for the NIR channel.}}
\label{Tab:Obs}
\end{table*}

\subsection{{\it XMM-Newton} observations}

{\em XMM-Newton} time was granted through a director's discretionary time observation (proposal ID~85338, P.I. J.~Sanz-Forcada) to observe WASP-76 system on 1 January 2020 with an overall 17\,ks exposure time. 
The target was not detected in any of the EPIC detectors. 
Given the distance of the object, we set an upper limit of $L_{\rm X}=2 ~ 10^{28}$\,erg\,s$^{-1}$ on the stellar X-ray luminosity, which implies a value of $\log L_{\rm X}/L_{\rm bol}<-6$ for
WASP-76.
This value indicates that the star has very low activity level. 
If detected, the spatial resolution would not allow us to separate a possible contribution from WASP-76B.


\section{Analysis} \label{sec:analysis}

\subsection{Atmospheric analysis}

The observations obtained in both the VIS and NIR channels were first corrected of H$_2$O and O$_2$ telluric absorption contamination using the {\tt Molecfit} tool \citep{Molecfit1,Molecfit2}. The extraction of the transmission spectrum around the different spectral lines in the VIS and NIR channels of CARMENES was performed following the methodology presented by \citet{Wytt2015} and \citet{2017CasasayasB}. Details on the extraction of the transmission spectrum around particular lines are presented in the following sections. 

In our analysis, we adopted the system and stellar parameters for WASP-76b summarised in Table~\ref{tab:params}. 
In the stellar rest frame, we measured a red-shift of the stellar lines of around $\sim 0.03$\,{\AA} in both VIS and NIR, which corresponds to 0.80--1.15\,km\,s$^{-1}$ in the regions studied here. Nevertheless, this wavelength shift is stable during the night and below the sampling of the spectrograph, so it does not affect the results. The systematic radial velocity $\gamma$ measured by \citet{West2016} produced the smallest wavelength shift in comparison with the other values from the literature \citep[e.g.][]{Ehrenreich2020}. 

\begin{table}[]
\centering
\caption{Physical and orbital parameters of the WASP-76b system$^a$.}
\begin{tabular}{lr}
\hline \hline
\\[-1em]
 Parameter  & Value\\ \hline
 \\[-1em]
 \multicolumn{2}{c}{\it Stellar parameters}\\\noalign{\smallskip}
   \\[-1em]
\quad  $T_{\rm eff}$ [K] & $6329\pm 65$ \\
  \\[-1em]
\quad  $\log g$ [cgs]& $4.196\pm 0.106$ \\
  \\[-1em]
\quad  [Fe/H] & $0.366\pm 0.053$ \\
  \\[-1em]
\quad $M_{\star}$ [$\rm{M_{\odot}}$]& $1.458\pm0.021$ \\
  \\[-1em]
\quad $R_{\star}$ [$\rm{R_{\odot}}$]& $1.756\pm0.071$ \\
  \\[-1em]
 \quad  $v\sin i_{\star}$ [km\,s$^{-1}$]& $1.48\pm0.28$\\
  \\[-1em]
 \multicolumn{2}{c}{\it Planet parameters}\\\noalign{\smallskip}
   \\[-1em]
 \quad $M_{\rm p}$ [$\rm{M_{Jup}}$]&  $0.894^{+0.014}_{-0.013}$\\
  \\[-1em]
 \quad $R_{\rm p}$ [$\rm{R_{Jup}}$]& $1.854^{+0.077}_{-0.076}$\\
 \\[-1em]
\quad $K_{\rm p}$ [km\,s$^{-1}$]& $196.52\pm0.94$ \\
  \\[-1em]
 \multicolumn{2}{c}{\it Transit parameters}\\\noalign{\smallskip}
   \\[-1em]
 \quad $T_{\rm 0}$ [BJD] & $2458080.626165^{+0.000418}_{-0.000367}$ \\
  \\[-1em]
\\[-1em]
 \quad $P$ [d] & $1.80988198^{+0.00000064}_{-0.00000056}$ \\
  \\[-1em]
 \quad $T_{14}$ $^b$ [min] & $230$\\
  \\[-1em]
 \quad $T_{12}$ $^b$ [min] & $23.6$\\
 \\[-1em]
 \multicolumn{2}{c}{\it System parameters}\\\noalign{\smallskip}
   \\[-1em]
  \quad $a/R_{\star}$& $4.08^{+0.02}_{-0.06}$ \\
 \\[-1em]
 \quad $i_{\rm p}$ [deg]& $89.623^{+0.005}_{-0.034}$\\
 \\[-1em]
 \quad $e$& 0 (fixed)\\
\\[-1em]
\quad $K_{\star}$ [m\,s$^{-1}$]& $116.02^{+1.29}_{-1.35}$\\
  \\[-1em]
\quad $\gamma$ [km\,s$^{-1}$]& $-1.0733\pm0.0002$\\
  \\[-1em]
 \quad $\lambda$ [deg]& $61.28^{+7.61}_{-5.06}$\\
\\[-1em]
\lasthline
\end{tabular}
\tablefoot{\tablefoottext{a}All values are adopted from \citet{Ehrenreich2020}, except the system velocity ($\gamma$), which is taken from \citet{West2016}. \tablefoottext{b}$T_{14}$ is the total transit duration, between the first and fourth contacts of the transit, while $T_{12}$ is the duration of the ingress, between the first and second contacts of the transit.}
\label{tab:params}
\end{table}

\subsubsection{Transmission spectrum in the VIS channel} \label{subsec:vis}

Using the observations in the VIS channel of CARMENES, we analysed the following spectral lines (all wavelengths given in vacuum): the \ion{Na}{i} doublet at $5891.58$\,{\AA} and $5897.56$\,{\AA}, H$\alpha$ at $6564.60$\,{\AA}, \ion{Li}{i} at $6709.61$\,{\AA} and $6709.76$\,{\AA}, the \ion{K}{i} line at $7701.08$\,{\AA}, and the \ion{Ca}{ii} IRT triplet at $8500.35$\,{\AA}, $8544.44$\,{\AA}, and $8664.52$\,{\AA}. In summary, to extract the transmission spectrum of the exoplanet, we shifted the observed spectra to the stellar rest frame using the system velocity ($\gamma$), the stellar radial-velocity semi-amplitude ($K_{\star}$; see values in Table~\ref{tab:params}), and the barycentric radial-velocity values calculated with {\tt caracal}. Then, we combined all the out-of-transit spectra to built a high S/N master stellar spectrum, which we used to remove the stellar contribution from the observations by computing the ratio of each individual spectrum to this master spectrum. The resulting residuals were shifted to the planet rest frame using the radial-velocity semi-amplitude $K_{\rm p} = 196.52$\,km\,s$^{-1}$ \citep{Ehrenreich2020}. Finally, the in-transit exposures between the first and fourth contacts of the transit were combined to obtain the transmission spectrum of the planet. Due to the low S/N in the stellar line cores, this combination was performed using the weighted mean with weights $1/\sigma_i^2$, with $\sigma_i$ being the uncertainties of each pixel resulting from the photon noise propagation \citep{Wytt2015}. Further, due to the difference in S/N between the two observing nights, the final transmission spectrum was obtained as the combination of the two individual transmission spectra using the mean nightly S/N as weights ($w_i = {\rm \overline{S/N}}_i^2/\sum_i{\rm \overline{S/N}}_i^2$, where ${\rm \overline{S/N}_i}$ is the mean S/N of the night $i$). This weighting results in a contribution of $\sim70\,\%$ and $\sim30\,\%$ for the first and second night, respectively.

A particular situation was observed for the \ion{Na}{i} lines, which show strong telluric \ion{Na}{i} emission and low S/N in the centre of the stellar line cores (S/N $\sim$ 5--20). Although the telluric \ion{Na}{i} emission was monitored with fibre B, the difference in efficiency between the two fibres, the stellar light contamination in fibre B, and the fact that this emission is not located in the continuum but inside the stellar \ion{Na}{i} lines made an accurate correction especially challenging. For this reason, we extracted the transmission spectrum using three different approximations: 
($i$) ignoring the presence of the contamination and the noise in the core of the lines, 
($ii$) attempting a correction of the telluric \ion{Na}{i} emission with fibre B observations following the methodology described in Section~\ref{subsec:nir} for the OH lines around \ion{He}{i}, and 
($iii$) masking a region of $\pm10$\,km\,s$^{-1}$ around the core of each \ion{Na}{i} line before combining the in-transit residuals to avoid the propagation of this contamination to the final transmission spectrum (see Fig.~\ref{fig:ts_NaI} for clarity). For the telluric correction case, as the \ion{Na}{i} emission lines in fibre A are located inside the stellar \ion{Na}{i} lines, we used five other isolated telluric emission lines in the VIS channel to measure the scaling factor between fibres. Our estimation for this scaling factor was $\sim0.8$ for both nights (see the following section for more details).

\subsubsection{Transmission spectrum around \ion{He}{i}} \label{subsec:nir}

\begin{figure}[]
\centering
\includegraphics[width=0.5\textwidth]{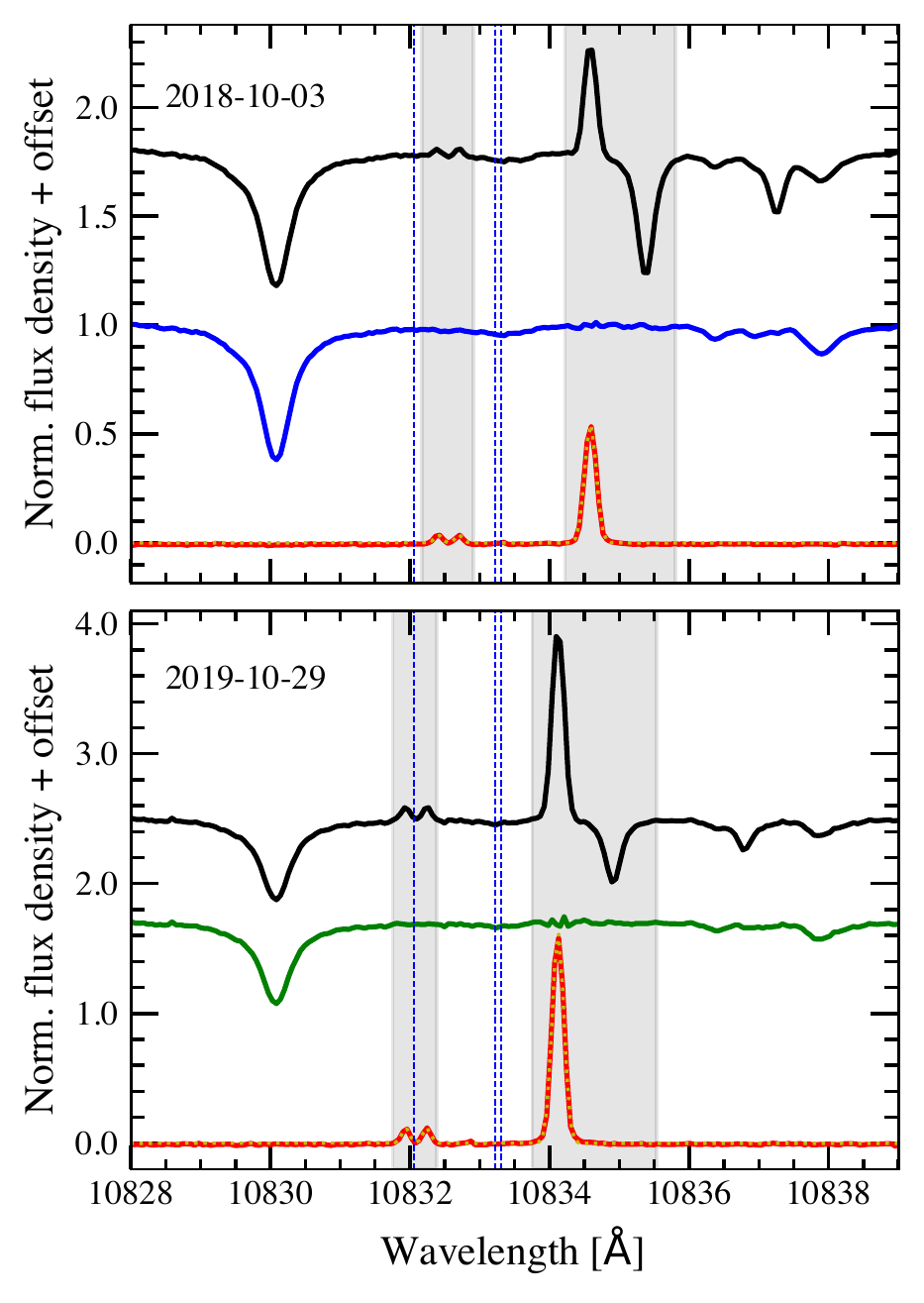}
\caption{Combined out-of-transit observations around the \ion{He}{i} triplet lines for the first ({\it top panel}) and second ({\it bottom panel}) nights in the stellar rest frame. The observed stellar spectrum (fibre A) is shown in black and the telluric-absoption-corrected spectrum in blue (first night) and green (second night). The sky spectrum (fibre B) is presented in red, and the best fit profile in yellow dots. In light grey we indicate the regions that were masked due to telluric contamination (see details in Section~\ref{subsec:nir}), and the vertical blue-dashed lines mark the positions of the \ion{He}{i} triplet lines.}
\label{fig:HeI_star}
\end{figure}

We used the observations in the NIR channel to search for the \ion{He}{i} $\lambda$10830\,{\AA} triplet 
in the atmosphere of WASP-76b. Surrounding the three \ion{He}{i} lines, we found water absorption and OH emission lines from the Earth's atmosphere. 
The observations of both nights were planned at dates with a favourable barycentric velocity of the Earth, to avoid a complete overlap of the telluric emission contamination with a possible exoplanet \ion{He}{i} signal. However, the emission lines remained close to the \ion{He}{i} lines and needed to be treated carefully. The impact of the telluric contamination can be seen in Fig.~\ref{fig:HeI_star}.

Here, we dealt with the telluric lines following two different approaches (see also \citealt{Palle2020a}), and extracted the transmission spectra in both cases. In the first case, we masked the emission and absorption features (see masked regions in
Fig.~\ref{fig:HeI_star}). Those masked pixels were not considered when computing the transmission spectrum, meaning that some of the points in the transmission spectrum are represented by less pixels than the remaining points and, therefore, show larger uncertainties (see details in Section~\ref{sec:results}).

\begin{figure*}[]
\centering
\includegraphics[width=1\textwidth]{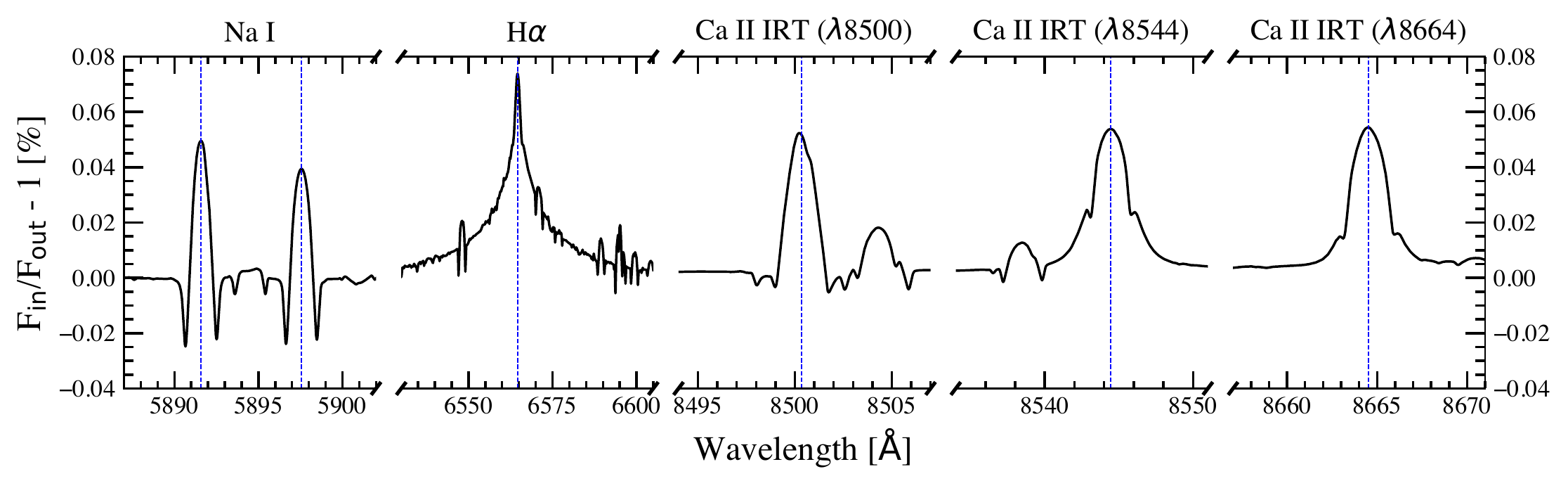}
\caption{Modelled RM and CLV effects on the transmission spectrum of WASP-76b around some of the spectral lines studied here: \ion{Na}{i} doublet, H$\alpha$, and \ion{Ca}{ii} IRT lines. These effects are calculated based on the ATLAS9 stellar atmospheric models with LTE. The transmission spectra are calculated in the planet rest frame. The vertical blue-dashed lines show the laboratory position of the spectral lines in vacuum. }
\label{fig:models}
\end{figure*}

The second method involved the correction of both telluric absorption and emission lines. The water absorption was corrected with {\tt Molecfit}. The OH emission features were corrected with the sky spectra from fibre B. The information of both fibres was extracted following the same methodology \citep[FOX,][]{Zechmeister2014}., but their efficiency is not identical and, thus, a scaling factor between the two fibres was required. To determine it, we computed a master sky spectrum using fibre B and a master stellar spectrum using fibre A, both computed considering only out-of-transit data. Then, we found the best scaling factor between the two master spectra, resulting in scaling factors of $\sim 0.89$ and $\sim 0.92$ for the first and second nights, respectively, which were assumed to be constant during the nights.

To correct the emission lines in each individual spectrum, we computed a synthetic emission spectrum fitting three Gaussian profiles (one per emission line) to each individual sky spectrum (see Fig.~\ref{fig:HeI_star}), so we accounted for the airmass change. Then, this sky model was scaled by the factor previously derived for the particular night, and subtracted from the observed stellar spectrum. Subtracting this synthetic sky spectrum instead of the sky observations directly prevented us from adding additional noise to the stellar spectrum. Once the telluric emission contamination was corrected, we extracted the transmission spectrum in the vicinity of \ion{He}{i} following the process detailed in Section~\ref{subsec:vis}.

\subsection{The Rossiter-McLaughlin and centre-to-limb variation effects on the transmission spectrum of WASP-76b}

The impact of the Rossiter-McLaughlin (RM) and centre-to-limb variation (CLV) effects on the transmission spectrum of WASP-76b was estimated assuming the stellar and planetary parameters derived by \citet{Ehrenreich2020}, and the methodology presented by \citet{Yan2017A&A...603A..73Y} and \citet{YanKELT9}. This same method has been used to estimate the strength of these effects in several recent atmospheric studies performed using high-resolution spectroscopy such as \citet{Borsa2020}, \citet{Casasayas2020}, and \citet{Chen2020}, among others. Here, in particular, we use the ATLAS9 stellar models \citep{ATLAS92003}, assuming local thermodynamic equilibrium (LTE) and solar abundance. 

WASP-76b is on a misaligned orbit, with an obliquity of $\lambda=61.28$\,deg, and crosses the stellar disc with an impact parameter of $b=0.027$ \citep{Ehrenreich2020}. The star rotates slowly, with a projected rotational velocity $v\sin i_{\star} \approx$ 1.48\,km\,s$^{-1}$. As discussed in previous atmospheric studies of this planet (\citealt{Seidel2019}, \citealt{Ehrenreich2020}, and \citealt{Tabernero2020}), the impact of the RM and CLV effects on the transmission spectrum of this planet is weak due to the rapid motion of the planet ($K_{\rm p} \approx 196.52$\,km\,s$^{-1}$) in comparison with the projected radial velocities of the stellar surface that the planet blocks during the transit. For this reason, when shifting the spectra to the rest frame of the planet, the RM and CLV effects are diluted over a wide range of wavelengths. 

In Fig.~\ref{fig:models} we show the predicted RM and CLV effects on the transmission spectrum of WASP-76b, around the strongest spectral lines studied here. At the centre of the \ion{Ca}{ii} IRT and \ion{Na}{i} lines, for example, the impact of these effects on the transmission spectrum is $\sim 0.05\,\%$ in relative flux. H$\alpha$ shows a slightly larger effect ($\sim 0.07\,\%$), while the impact on the \ion{Li}{i} and \ion{K}{i} lines is weaker. The transmission spectra derived in this work show a typical uncertainty per pixel of $\sim0.2\,\%$. Thus, in the analysis presented here, the impact of these effects on the transmission spectrum is not significant, as the corrections remain below the precision achieved in the observations.

The effects for the \ion{He}{i} lines formed in the stellar chromosphere cannot be modelled following the same methodology, as these stellar models are limited to photospheric lines. However, the stellar \ion{He}{i} line is weak (see Fig.~\ref{fig:HeI_star}). \citet{Nortmann2018Science} simulated the RM effect on the \ion{He}{i} lines of \object{WASP-69}b using the observed stellar spectrum, and found a strength of $\sim 0.07\,\%$ in the final transmission spectrum when assuming an increased radius of $1.73~R_{\rm p}$ due to the \ion{He}{i} envelope. Taking into account the semi-amplitude of the RM radial velocities, the overlap of the RM and planet radial velocities, and the strong \ion{He}{i} lines in the stellar spectrum of WASP-69 in comparison with \object{WASP-76}, we concluded that the impact of the RM effect on the \ion{He}{i} lines is negligible for WASP-76b. {The CLV, on the other hand, could not be estimated using the observed (full disc integrated) stellar spectrum, as the computation of stellar spectra at different limb-darkening angles is necessary.}

\subsection{Activity} \label{sec:activity}

Stellar activity and the presence of star spots occulted by the exoplanet during the observations can influence the transmission spectrum. Here we study the variability of some activity indicators and search for possible star spot occultation using photometric light curves. 

\subsubsection{Activity indicators}

WASP-76 is an F7 dwarf, which has not shown any evidence of stellar activity in previous studies (e.g. \citealt{Zak2019}, \citealt{Essen2020}). However, some of the stellar lines studied here for atmospheric purposes may partially be produced in the chromosphere and, consequently, could be affected by stellar activity \citep{Barnes2017}. For example, \citet{Khalafinejad2018} studied the variations of the \ion{Ca}{ii} IRT lines during the transit of an exoplanet as an indicator of stellar activity. To monitor the stellar variability during our observations, we processed the {\tt Molecfit} telluric corrected data from the VIS channel using the Spectrum radial velocity analyser ({\tt serval}; \citealt{SERVAL}), which is the standard CARMENES pipeline to derive radial velocities as well as several activity indicators (see Section~\ref{sec:obs_carm}). In particular, we looked at the chromatic radial-velocity index (CRX),and the differential line width (dLW, details on the definition and calculation of these indicators can be found in the work by \citealt{SERVAL}).

\begin{figure}[]
\centering
\includegraphics[width=0.45\textwidth]{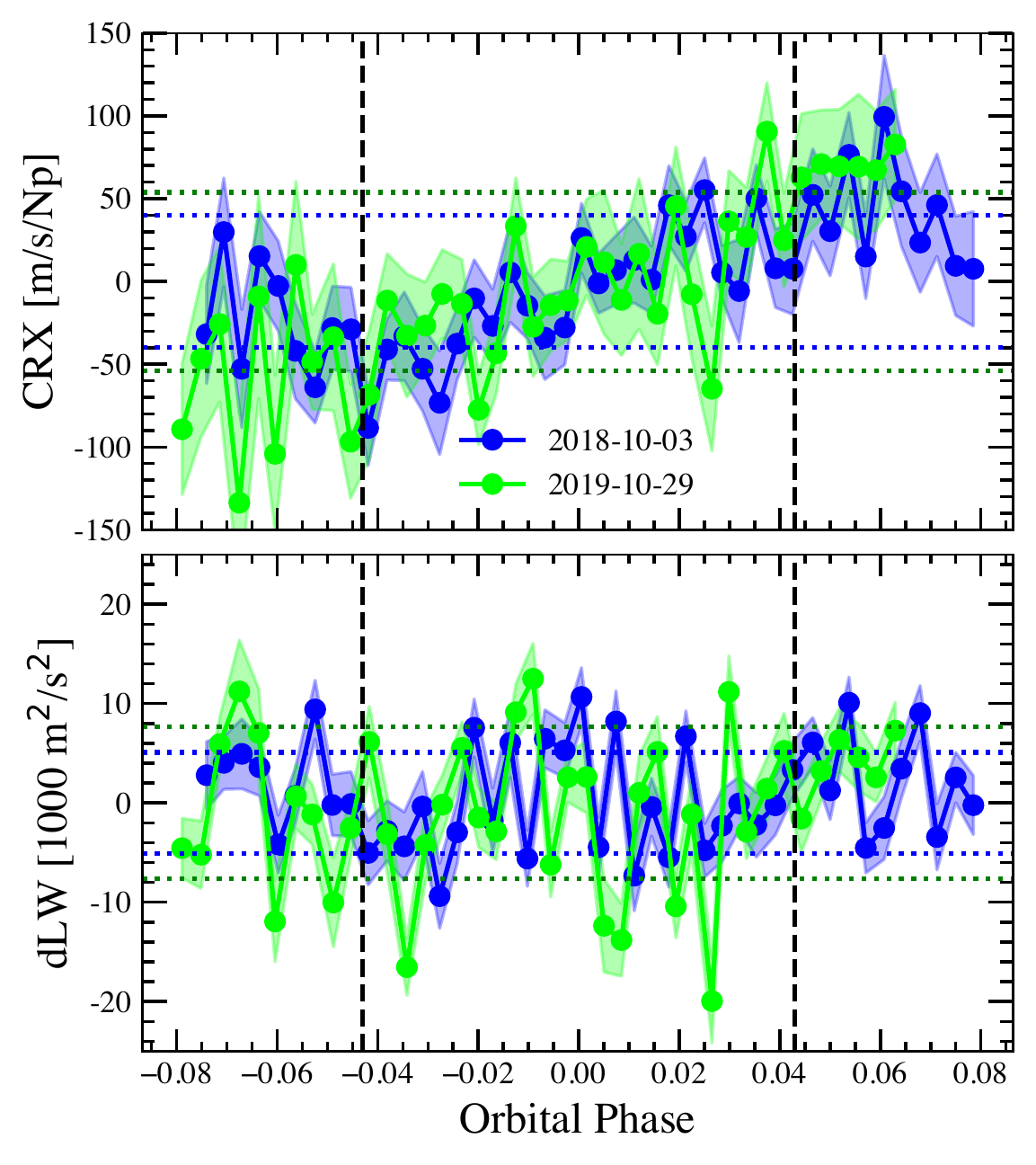}
\caption{Activity indices derived with {\tt serval} for the two observing nights, shown in blue and green lines, respectively. 
{\em Top panel}: chromatic radial-velocity index (CRX). 
{\em Bottom panel}: differential line width (dLW). The horizontal dotted lines show the standard deviation of the values for the first (blue) and second night (green).}
\label{fig:activ_index}
\end{figure}

In Fig.~\ref{fig:activ_index} we show the evolution of the CRX and dLW indices on the two nights of observation. None of the two indices showed evidence of significant variations produced by flares or other sources of stellar variability as shown by, for example, \citet{Palle2020b}. If we focus on smaller variations, we may see a systematic evolution of the CRX over time. This index is sensitive to spots, and observing a trend 
could be indicative of the movement of active regions. However, considering the typical rotation period of F stars ($\sim$15--40\,d; \citealt{SuarezM2015,SuarezM2016}), one would not expect this fast evolution of stellar spots in WASP-76 observations (see also Section~\ref{sec:spots}). Instead, the origin of this (weak) variation might be the combination of some telluric residuals or 
chromaticity induced by the presence of WASP-76B. 
In addition, we explored the core of the stellar H$\alpha$, \ion{Na}{i}, and \ion{Ca}{ii} IRT lines, but we observed no sign of emission produced by stellar activity.

\subsubsection{TESS light curve} \label{sec:spots}

\begin{figure}[]
\centering
\includegraphics[width=0.49\textwidth]{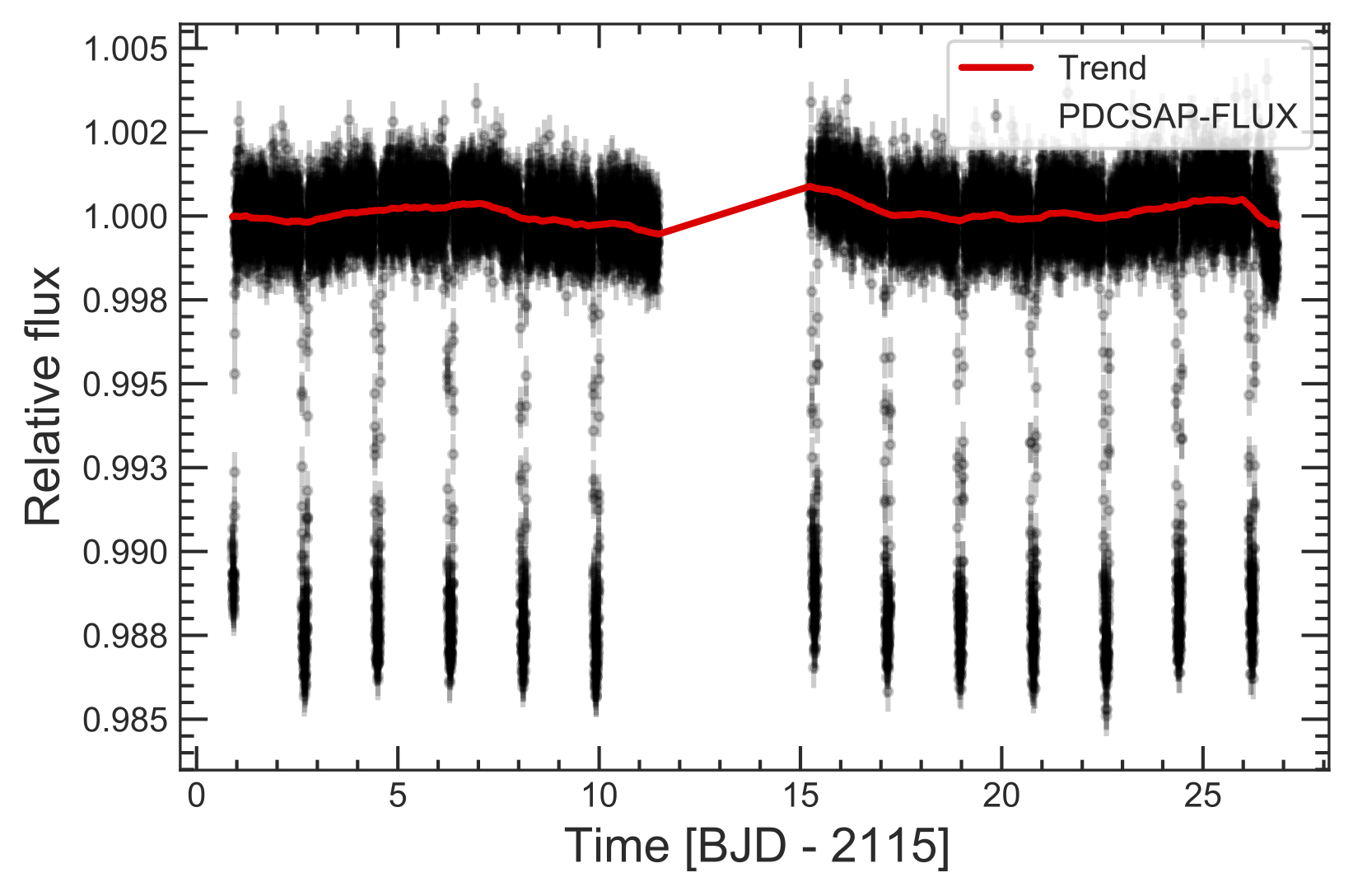}
\caption{{\em TESS} light curve (PDCSAP flux) of WASP-76 as observed during sector 30. The red line shows the trend obtained by applying a bi-weight detrending filter determined with \texttt{w{\={o}}tan}.}
\label{fig:PDC-WOTAN}
\end{figure}

WASP-76 was observed in 2-min short-cadence mode by the \textit{TESS} spacecraft in sector 30, using camera 1, between 2020-09-22 and 2020-10-21. We downloaded its corresponding light curve from the Mikulski Archive for Space Telescopes\footnote{\url{https://mast.stsci.edu}, \url{https://archive.stsci.edu/}} (MAST), produced by the Science Processing Operations Center at the NASA Ames Research Center \citep[SPOC;][]{SPOC}. SPOC provides simple aperture photometry (SAP) and systematic-corrected photometry as derived with the Pre-search Data Conditioning algorithm (PDCSAP). Since we are interested in analysing the transit light curve and also searching for stellar spot crossing events, we used the PDCSAP flux. Although the PDCSAP dataset had already been corrected for dominant systematics by default, we further corrected it for small systematics, which were still evident in the light curve. We performed a smooth de-trending of the light curve using \texttt{w{\={o}}tan} \citep{Hippke-19} with a window length of 1\,d and the bi-weight method. The PDCSAP light curve and the best trend from \texttt{w{\={o}}tan} are shown in Fig.~\ref{fig:PDC-WOTAN}. 

\begin{figure}[h]
\centering
\includegraphics[width=0.49\textwidth]{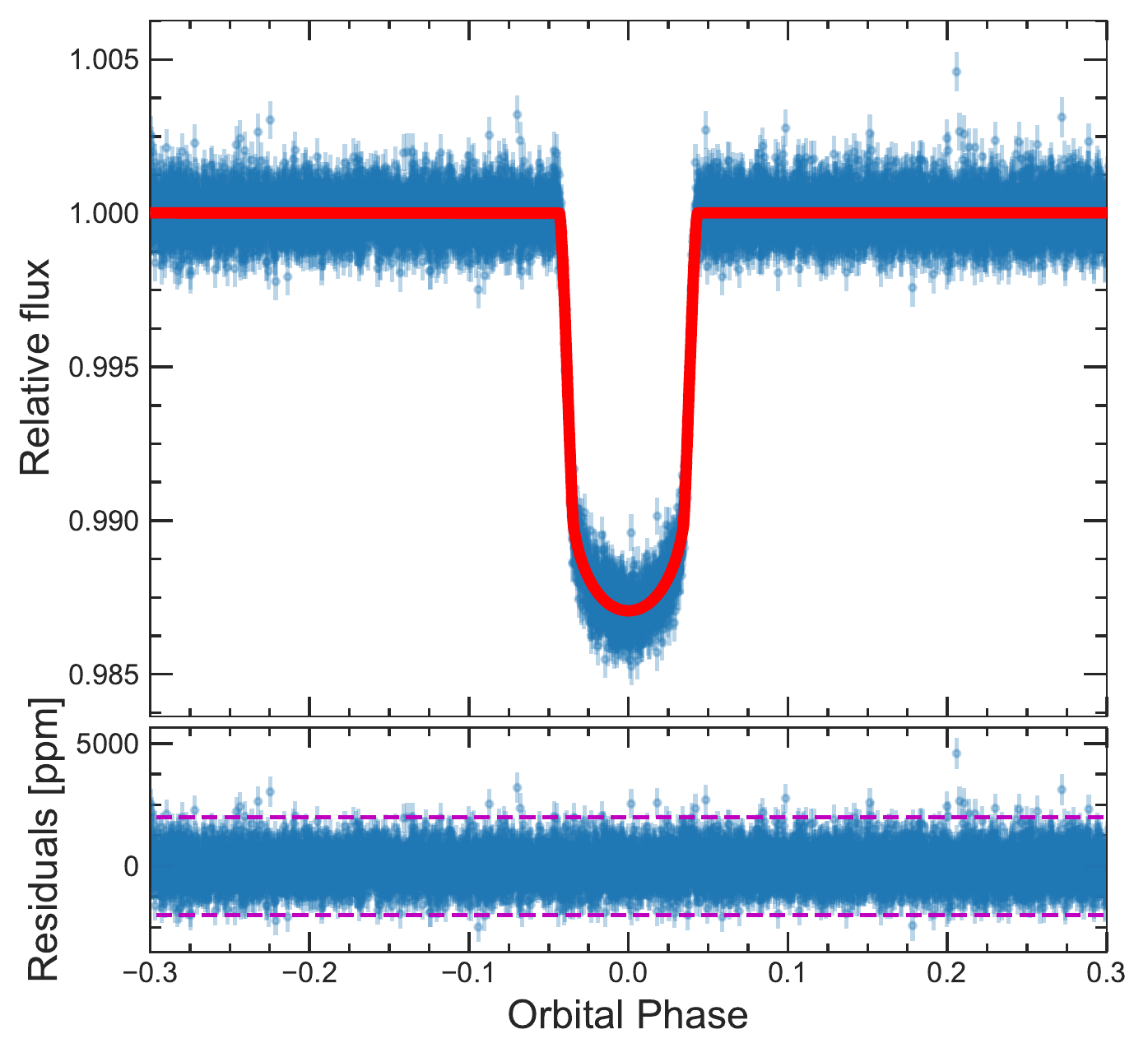}
    \caption{{\em Top panel}: TESS phase-folded light curve centred on the transit of WASP-76b. The red line shows the best-fit transit model. {\em Bottom panel}: residuals between the phase-folded light curve and the best-fit model. The magenta dashed lines depict the residual standard deviation multiplied by three, which points to no significant anomalies inside the transits in comparison to the points out of transit.}
\label{fig:PDC-FIT}
\end{figure}

We modelled the phase folded transit light curve with the \texttt{batman} package \citep{batman}. We only left the transit times (in case of transit timing variations), transit depth, and limb darkening coefficients as free parameters, and fixed the remaining parameters to the values reported in the literature. The best-fit parameters and associated uncertainties in our fitting procedure were derived using a Markov chain Monte Carlo (MCMC) analysis implemented in the \texttt{emcee} python package \citep{emcee}. Fig.~\ref{fig:PDC-FIT} shows the phase-folded transit light curve of WASP-76b and the best-fit transit model.

Several high-precision photometric studies have shown that when a transiting planet crosses stellar spots, anomalies can be generated in the transit light curves, which can lead to inaccurate estimates of the planetary parameters \citep[e.g.][]{Czesla-09, Nutzman-11, Sanchis-Ojeda-11a, Sanchis-Ojeda-11b, Oshagh-13, Ioannidis-16}. An inspection of the residuals between the phase-folded transit light curve of WASP-76b and the best-fit model, as displayed in Fig.~\ref{fig:PDC-FIT}, does not show the presence of extra noise inside the transits in comparison to the data out of transits, which could be attributed to spot crossing anomalies. For this reason, we concluded that spot crossing events during the CARMENES transit observations presented here are unlikely. Additional evidence of the low activity level of the star was observed with the Optical Monitor on board {\em XMM-Newton}, which showed stochastic variability $\leq 18\,\%$ in the $UVW2$ filter ($\lambda$1620--2620\,{\AA}) that we could not associate to the presence of faculae.

\subsection{The binary companion: WASP-76B} \label{sec:companion}

Due to the small separation of both stars of the system ($\rho \approx 0.44$\,arcsec), WASP-76B is indeed contaminating our spectra. However, the magnitude difference between both stars is $\Delta K = 2.30$\,mag, which means a flux ratio of only $\sim9$. Using ESPRESSO observations, \citet{Ehrenreich2020} performed an in depth study of the impact of this stellar companion in their spectra, as it resides in the limit of the $0.5$\,arcsec-radius of the ESPRESSO fibre \citep{Pepe2020} and, therefore, the contamination could change with the seeing. Nevertheless, they concluded that, although the seeing of their observations was $<0.7$\,arcsec, WASP-76B remained undetectable in their observations and discussed several scenarios that could hide its presence.

For our CARMENES observations, the size of the fibre on the sky is $1.5$\,arcsec and, consequently, we expected the presence of WASP-76B to be included in all observations, independently of the seeing. Thus, if existing, we expected this contamination to be stable during the observations and unlikely to affect our results. As seen by \citet{Ehrenreich2020}, no spectral features from WASP-76B are distinguished in our observations. This is further examined in Sect.~\ref{sec:results}.

\section{Results and discussion} \label{sec:results}


\subsection{Planetary atmospheric transmission in the \ion{Ca}{II} IRT lines}  \label{sec:results_Ca}

\subsubsection{Transmission spectrum around single lines}

\begin{figure*}[]
\centering
\includegraphics[width=1\textwidth]{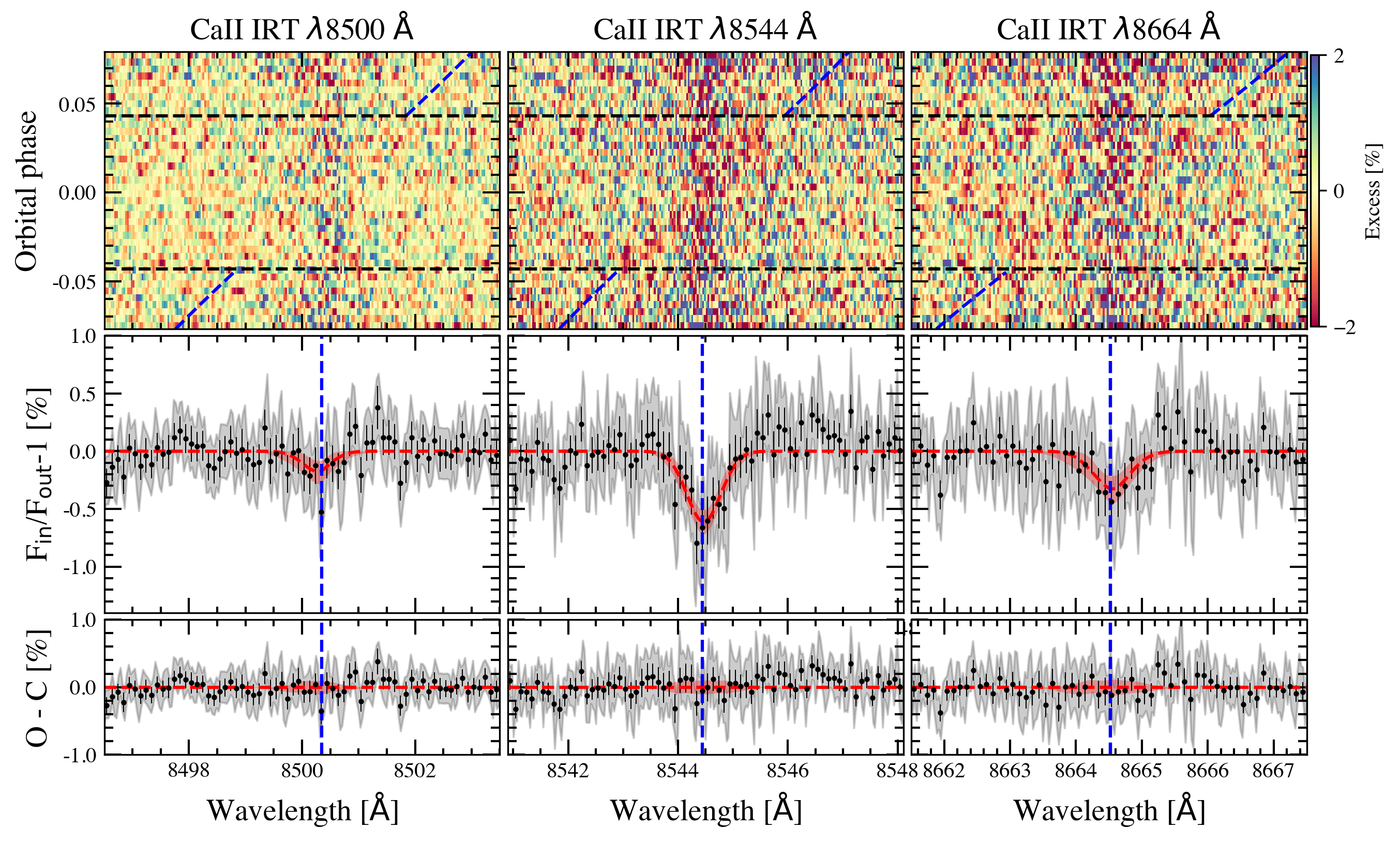}
\caption{Transmission spectra obtained around the three lines of the \ion{Ca}{ii} IRT triplet, as a result of combining the two nights. 
{\em Top panels}: tomography maps shown in the stellar rest frame. The orbital phase of the planet is presented in the vertical axis, the wavelength (in vacuum) in the horizontal axis, and the flux excess information is shown in the colour bar (in \%). The data are shown binned by $0.004$ in orbital phase. The tilted blue-dashed lines show the position of the spectral lines in the planet rest frame. The black-dashed lines show the first and last contact of the transit ($\phi=\pm0.043$; \citealt{Ehrenreich2020}). Middle panels: transmission spectra computed considering the data between the first and last contacts of the transit. In light grey we show the uncertainties of the original data sampling, and in black the data binned by $0.1$\,{\AA}. The blue-dashed lines show the laboratory wavelengths of the lines, the red-dashed line the best-fit Gaussian profile, and the light red regions indicate the 1$\sigma$ uncertainties of the profiles. Bottom panels: Residuals between the observed transmission spectra and the best-fit Gaussian profiles shown in the middle panel. }
\label{fig:TS_CaIRT}
\end{figure*}

\begin{figure}[]
\centering
\includegraphics[width=0.49\textwidth]{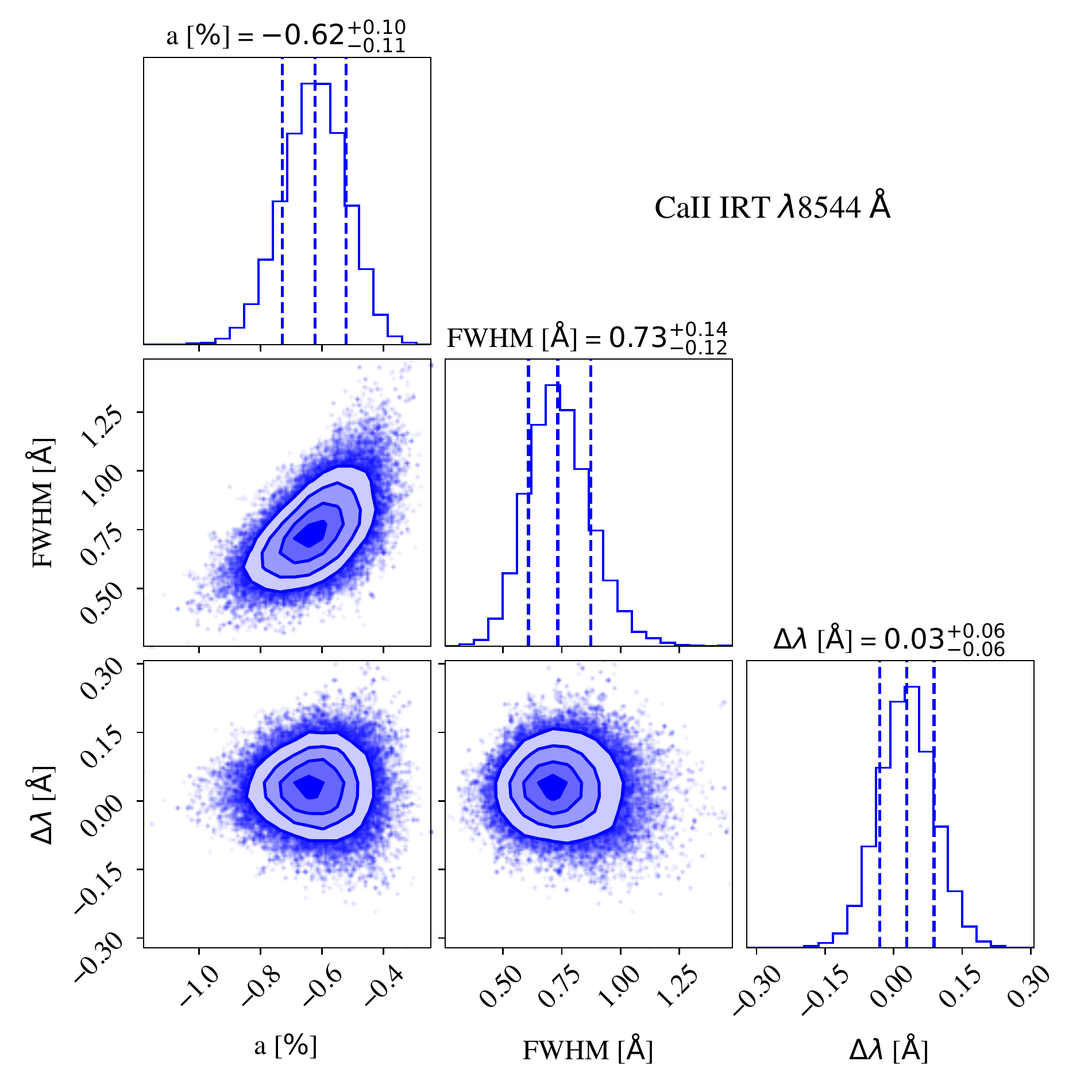}
\caption{MCMC probability distribution of the Gaussian profile parameters for the \ion{Ca}{ii} IRT $\lambda 8544$\,{\AA} absorption line observed in the CARMENES transmission spectrum of WASP-76b.}
\label{fig:MCMC_Ca2}
\end{figure}

When combining the two transit observations, we detect the individual \ion{Ca}{ii} IRT lines in the transmission spectrum of WASP-76b at $1.9\sigma$, $5.6\sigma$, and $2.9\sigma$ level of significance, respectively. The tomography maps and transmission spectra are presented in Fig.~\ref{fig:TS_CaIRT}. 
The faintest line at $8500$\,{\AA} cannot be distinguished from the noise level in the maps and is not significantly detected ($1.9\sigma$) even after combining the two nights. 
Absorption from the line at $8664$\,{\AA} is poorly recovered ($2.9\sigma$) and only slightly visible in the tomography maps. However, the absorption signal of the strongest line at $8544$\,{\AA} can be followed in the tomography maps and is detected after combining the two nights ($5.6\sigma$). 
The best-fit parameters of the Gaussian profiles describing the three lines are presented in Table~\ref{Tab:CaII_res}. These parameters were obtained with the MCMC algorithm implemented in {\tt emcee} \citep{emcee} using $10^4$ steps, $10$ walkers, and a range of $\pm300$\,km\,s$^{-1}$ surrounding the spectral lines in the transmission spectrum. None of the lines show a significant blue- or red-shift. The MCMC probability distribution of the \ion{Ca}{ii} IRT line at $8544$\,{\AA} is shown in Fig.~\ref{fig:MCMC_Ca2}. The distributions of the other lines are presented in Fig.~\ref{fig:corner}.

\begin{table}[]
\centering
\caption{Best-fit values of the Gaussian profiles fitted to the \ion{Ca}{ii} IRT lines of the transmission spectrum.}
\begin{tabular}{cccc}
\hline
\hline
\noalign{\smallskip}
 \ion{Ca}{ii} IRT & Contrast   &  FWHM   &  $\Delta v$ \\
         line [{\AA}]&  [$\%$] & [{\AA}]& [km\,s$^{-1}$]  \\
\noalign{\smallskip}
\hline 
\noalign{\smallskip}

$8500$ & $-0.22\pm0.11$ & $0.6\pm0.3$  & $-3\pm5$ \\

$8544$ & $-0.62\pm0.11$ & $0.7\pm0.1$  & $1\pm2$ \\

$8664$ & $-0.35\pm0.12$ & $0.7\pm0.3$  & $1\pm4$ \\

\noalign{\smallskip}
\hline
\end{tabular}\\
\label{Tab:CaII_res}
\end{table}

\begin{figure*}[]
\centering
\includegraphics[width=0.98\textwidth]{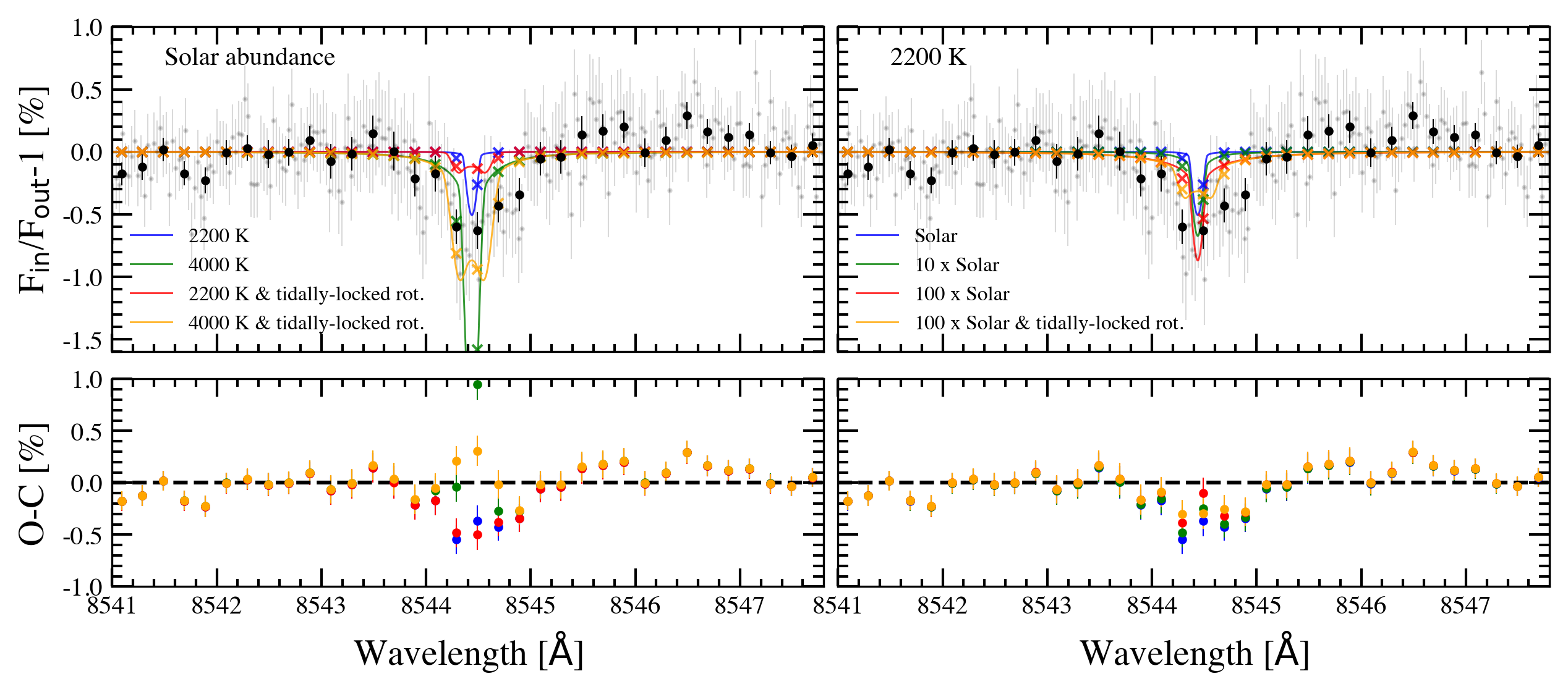}
\caption{CARMENES transmission spectrum of WASP-76b obtained around the \ion{Ca}{ii} IRT $\lambda8544$\,{\AA} line compared to different atmospheric models. {\em Top panels:} The original data are shown in light grey, the data binned by $0.2$\,{\AA} are shown in black dots, and the models binned by the same amount are presented in coloured crosses. {\em Top left panel}: models computed assuming solar abundance. The blue and green lines show the transmission models derived assuming $2200$\,K and $4000$\,K, without considering rotational broadening. In red and orange we present the atmospheric models at these same temperatures, but considering a tidally locked rotational broadening. {\em Top right panel}: atmospheric models computed assuming a temperature of $2200$\,K but considering different abundances: solar (blue; same as in the top panel), 10 times solar (green), 100 times the solar abundance without (red) and with (orange) tidally locked rotation. {\em Bottom panels}: residuals between the transmission spectrum and the different models (same colours as in top panels). The black-dashed line shows the reference at $0$\,{\%}.}
\label{fig:TS_CaIRT_mod}
\end{figure*}

To compare with the observed transmission spectrum, we calculated atmospheric models of the \ion{Ca}{ii} IRT lines as presented by \citet{Yan2019}. In summary, we used the {\tt petitRADTRANS} code \citep{petitRADTRANS2019} and the line list of \citet{Kurucz2011}. In this calculation we assumed that all the calcium is in the form of Ca$^+$ due to the high temperatures and low pressures where the \ion{Ca}{ii} lines are formed. The atmosphere was assumed to be isothermal, and the mean molecular weight $\mu$ equal to $1.3$. These models were calculated considering two different temperatures: 2200\,K (the equilibrium temperature of the exoplanet) and $4000$\,K, assuming that the exosphere where these lines originate is significantly hotter than the equilibrium temperature (e.g. \citealt{Yan2019} and \citealt{Turner2020}). 
We compared these non-rotating models to models in which the broadening due to the tidally locked rotation of the planet ($5.3$\,km\,s$^{-1}$; \citealt{Ehrenreich2020}) is considered. During the transit, the atmosphere of an exoplanet is observed as an annulus around the planet. In consequence, due to the tidally locked rotation, the atmosphere is red-shifted in the morning terminator and blue-shifted at the evening terminator, producing a double peak in the spectral lines of the transmission spectrum.
Finally, we computed a set of models with solar and super-solar abundances, since \citet{Tabernero2020} reported a stellar metallicity slightly larger than solar for WASP-76 and, in the gravitational-instability theory for giant-planet formation (e.g. \citealt{Boss1997}), giant planets are expected to end up with a metallicity similar to the host star. The comparison of these models with the observed transmission spectrum around \ion{Ca}{ii} IRT $\lambda 8544$\,{\AA}, which has the highest S/N, is presented in Fig.~\ref{fig:TS_CaIRT_mod}.

When computing the residuals between the models and the data, as illustrated by the bottom panel of Fig.~\ref{fig:TS_CaIRT_mod}, the S/Ns of the observations are too low to clearly differentiate between models. However, simple $\chi ^2$ calculations suggest that the observed transmission spectrum is better described by the model including tidally locked rotational broadening, a temperature of $4000$\,K, and solar abundance (see $\chi ^2$ values in Table~\ref{tab:chi2}).
The precision achieved in the transmission spectrum is not sufficient to confirm or discard the double peak in the lines core that could be produced by the rotation scenario, or to confirm a blueshift induced by a day-to-night side wind \citep{Ehrenreich2020}. The additional broadening of the lines could be attributable to vertical winds, as derived by \citet{Seidel2021} using ESPRESSO observations of the same planet, or produced by turbulence in the exoplanet exosphere \citep{Jager1954}. Another source of broadening is the radial-velocity change of the planet during an exposure. In $492$\,s (Table~\ref{Tab:Obs}) the planet moves about 4\,km\,s$^{-1}$ ($\sim 0.11$\,{\AA}), that is $\sim$3--4 pixels around the \ion{Ca}{ii} IRT lines. Thus, the FWHMs measured for these lines are more than five times larger than this value and, consequently, its contribution to the line broadening is relatively small and should not affect our conclusions. 

\begin{table}[]
\centering
\caption{Summary of the $\chi ^2$ values for the different models$^{a}$.}
\begin{tabular}{llcr}
\hline
\hline
\noalign{\smallskip}
Temperature & Abundance & Tidally locked & $\chi ^{2}$\\

 & & broadening? & \\

\noalign{\smallskip}
\hline 
\noalign{\smallskip}

2200\,K  &  solar & no & 60\\
         &  solar & yes & 59\\
         &  10 $\times$ solar & no & 51\\
         &  100 $\times$ solar & no & 39\\
         &  100 $\times$ solar & yes & 36\\
4000\,K  &  solar & no & 101\\
         &  solar & yes & 34\\

\noalign{\smallskip}
\hline
\end{tabular}\\
\tablefoot{\tablefoottext{a} Calculated in a region of $\pm30$\,km\,s$^{-1}$ surrounding the \ion{Ca}{ii} IRT $\lambda8544$\,{\AA} line in the transmission spectrum.}
\label{tab:chi2}
\end{table}

The detection of the \ion{Ca}{ii} IRT lines in the transmission spectrum of WASP-76b was expected, as ionised calcium was already detected in the atmosphere of this same planet by \citet{Tabernero2020} using ESPRESSO observations, but in this case through the \ion{Ca}{ii} H\&K lines. \citet{Tabernero2020} found an average line depth of $2.4\,\%$, FWHM ranging from $\sim$ 21--78\,km\,s$^{-1}$ depending on the night and line, and a shift of the lines consistent with zero. Here, as observed by \citet{Yan2019} for WASP-33b and KELT-9b, the absorption measured in the \ion{Ca}{ii} IRT lines is weaker than in the \ion{Ca}{ii} H\&K lines, and also do not show a significant shift. 
We also measured a FWHM of $\sim 25$\,km\,s$^{-1}$, in line with the other spectral lines detected by \citet{Tabernero2020}. To date, Ca$^+$ has been detected in five ultra-hot Jupiters: WASP-33b \citep{Yan2019}, KELT-9b (\citealt{Yan2019} and \citealt{Turner2020}), MASCARA-2b (\citealt{Casasayas2019} and \citealt{Nugroho2020}), WASP-76b \citep{Tabernero2020}, and WASP-121b \citep{Borsa2020}. Most of these studies have shown significant detections of Ca$^+$, while Ca has only been detected in WASP-121b \citep{Hoeijmakers2020_wasp121b}, showing consistency with the prediction that, at high temperatures, Ca is more abundant in its ionised state than in its neutral counterpart \citep{Helling2019}. However, this could also mean that Ca line lists are more limited than those for Ca$^+$, as discussed in \citet{Helling2021a}.

 In ultra-hot Jupiters, due to their tidally locked rotation and their proximity to the host star, the day and night-side atmospheres are expected to be significantly different in terms of composition \citep{Helling2019,Helling2021b,Helling2021a}. Transmission spectroscopy probes the atmosphere of the terminator region that, in ultra-hot Jupiters, has been observed to be less affected by disequilibrium processes than the two hemispheres, leading to a different atmospheric regime \citep{Molaverdikhani2020}. The detection of ionised atoms at the terminator could be due to the presence of strong transport mechanisms that move the atmospheric species from the day-side ionosphere to the terminator, or due to thermal ionisation in the upper exosphere of these planets \citep{Yan2019}. For the particular case of WASP-76b, \citet{Ehrenreich2020} detected asymmetric \ion{Fe}{i} absorption during the transit, explained as the combination of day-to-night winds and the lack of absorption from the nightside close to the morning (night-to-day) terminator due to the condensation of iron on the nightside \citep{Wardenier2021}. On the other hand, the \ion{Ca}{ii} H\&K lines measured by \citet{Tabernero2020} and the \ion{Ca}{ii} IRT lines measured by us are not significantly blue-shifted in comparison to other detected lines. This difference could be explained if, unlike the Fe, Ca$^+$ is formed at high altitudes of the atmosphere due to the high irradiation and it is present at both terminators.

\subsubsection{Template cross-correlation analysis of \ion{Ca}{ii}}

\begin{figure*}[]
\centering
\includegraphics[width=1\textwidth]{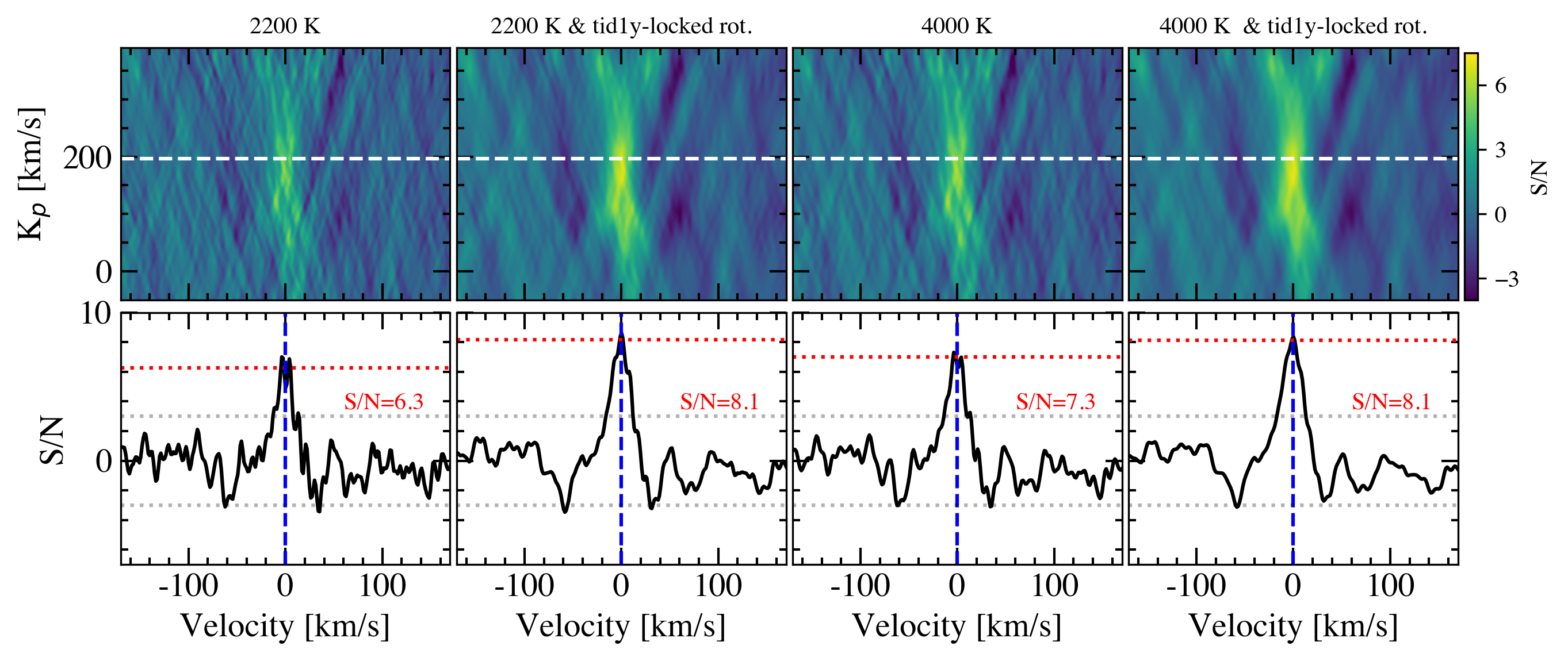}
\caption{\ion{Ca}{ii} IRT cross-correlation results using several atmospheric models computed under different assumptions (each column).
{\em Top panels}: $K_{\rm p}$-velocity map. The colour bar indicates the significance (S/N) of the measurement. The white-dashed lines indicate the expected $K_{\rm p} = 196.52$\,km\,s$^{-1}$. 
{\em Bottom panels}: cross-correlation result at the predicted $K_{\rm p}$ value. The blue-dashed lines show the 0\,km\,s$^{-1}$ radial velocity and the grey-dotted lines the $\pm3\sigma$ significance level. The red-dotted horizontal lines show the significance of the peak, obtained by fitting a Gaussian profile to the transmission signal.}
\label{fig:CC_CaIRT}
\end{figure*}

To gain S/N by searching for all three \ion{Ca}{ii} IRT lines, the atmospheric models presented in the previous section were cross-correlated with the observations to search for the three lines of the \ion{Ca}{ii} triplet at the same time and, thus, increase the S/N of the detected signal. We followed the methodology presented by \citet{Stangret2020}. The cross-correlation was applied to the three CARMENES orders containing the \ion{Ca}{ii} triplet lines (70, 71, 72), using steps of $1.3$\,km\,s$^{-1}$ (mean CARMENES resolution), and exploring a range of $\pm200$\,km\,s$^{-1}$. As illustrated by Fig.~\ref{fig:CC_CaIRT}, 
the \ion{Ca}{ii} triplet is detected independently of the model used in the cross-correlation, achieving a similar significance (S/N~$\sim$ 6--8), probably because two of the three lines of the triplet are faint (see Fig.~\ref{fig:TS_CaIRT}), and the differences between models are not significant for these particular observations (see Fig.~\ref{fig:TS_CaIRT_mod}). The results for the two individual nights are shown in Fig.~\ref{fig:cc_Ca_ind}. 

To be sure that the observed signal has a planetary origin, we computed the $K_{\rm p}$ maps by shifting the cross-correlation results in a range of $K_{\rm p}$ values from $-50$\,km\,s$^{-1}$ to $400$\,km\,s$^{-1}$, and combining the in-transit results for each $K_{\rm p}$ value (see the second row of Fig.~\ref{fig:CC_CaIRT}). The strongest correlation was found at the expected $K_{\rm p} \sim 196$\,km\,s$^{-1}$, confirming our findings around the single lines. The significance of the signals was calculated by dividing the combined results by the standard deviation computed in the regions between $-75$ and $-150$\,km\,s$^{-1}$, and $75$ and $150$\,km\,s$^{-1}$ \citep{birkby2017,SanchezLopez2019}. We rejected the possibility of these lines being the result of stellar spot occultation or contamination from the stellar companion WASP-76B because both individual transits showed consistent results at the expected $K_{\rm p}$ value. Occulting a stellar spot at the same latitudes in different epochs or finding contamination from WASP-76B at this particular $K_{\rm p}$ value is highly unlikely. As presented by \citet{Kesseli2021}, we decompose the \ion{Ca}{ii} cross-correlation signal considering only the first and only the second half of the transit (see also \citealt{Wardenier2021}). However, we did not find a significant difference in terms of radial-velocity and strength between the two signals that could be attributed to the asymmetry and condensation observed in \ion{Fe}{i} by \citet{Ehrenreich2020}.

\subsection{\ion{He}{i} absorption signal} \label{sec:results_HeI}

The \ion{He}{i} triplet lines at 10830\,{\AA} are tracers for extended atmospheres and atmospheric evaporation \citep{SeagerSas2000}, a process where the intense X-ray ($\sim$ 0.5--10\,nm) and extreme ultraviolet (EUV, 10--92\,nm) irradiation from the host star causes the atmosphere of hot gas planets to continuously expand resulting in mass loss \citep{Salz2016}. Photons with wavelengths $\lambda < 504$\,{\AA} can ionise He, producing He$^+$, which can populate the metastable $2^3S$ state of helium via the so-called photoionisation-recombination mechanism \citep{SanzForcada2008}. In a planetary system, XUV (X-ray and EUV) irradiation from the host star is, thus, expected to trigger the \ion{He}{i} triplet in the exoplanet atmosphere. One of the main advantages of using the near-infrared \ion{He}{i} lines is that they are not affected by interstellar medium \citep{Indriolo2009} and geocoronal absorption, unlike Ly$\alpha$ \citep{Oklop2018}, and are much less influenced by stellar activity \citep{Cauley2018}. Helium was detected for the first time in an exoplanet exosphere by \citet{Spake2018}, using {\em Hubble Space Telescope} observations, followed by the first ground-based detection by \citet{Nortmann2018Science} and \citet{Allart2018}, and the subsequent findings in other exoplanets (e.g. \citealt{Salz2018He,Allart2019,AlonsoFloriano2019HD209,Palle2020a}). These results have been used recently for constraining the mass loss rates of exoplanets suffering hydrodynamics escapes and determining their escape regime \citep{Lampon2020, Lampon2021,Lampon2021a}.

\begin{figure*}[]
\centering
\includegraphics[width=0.98\textwidth]{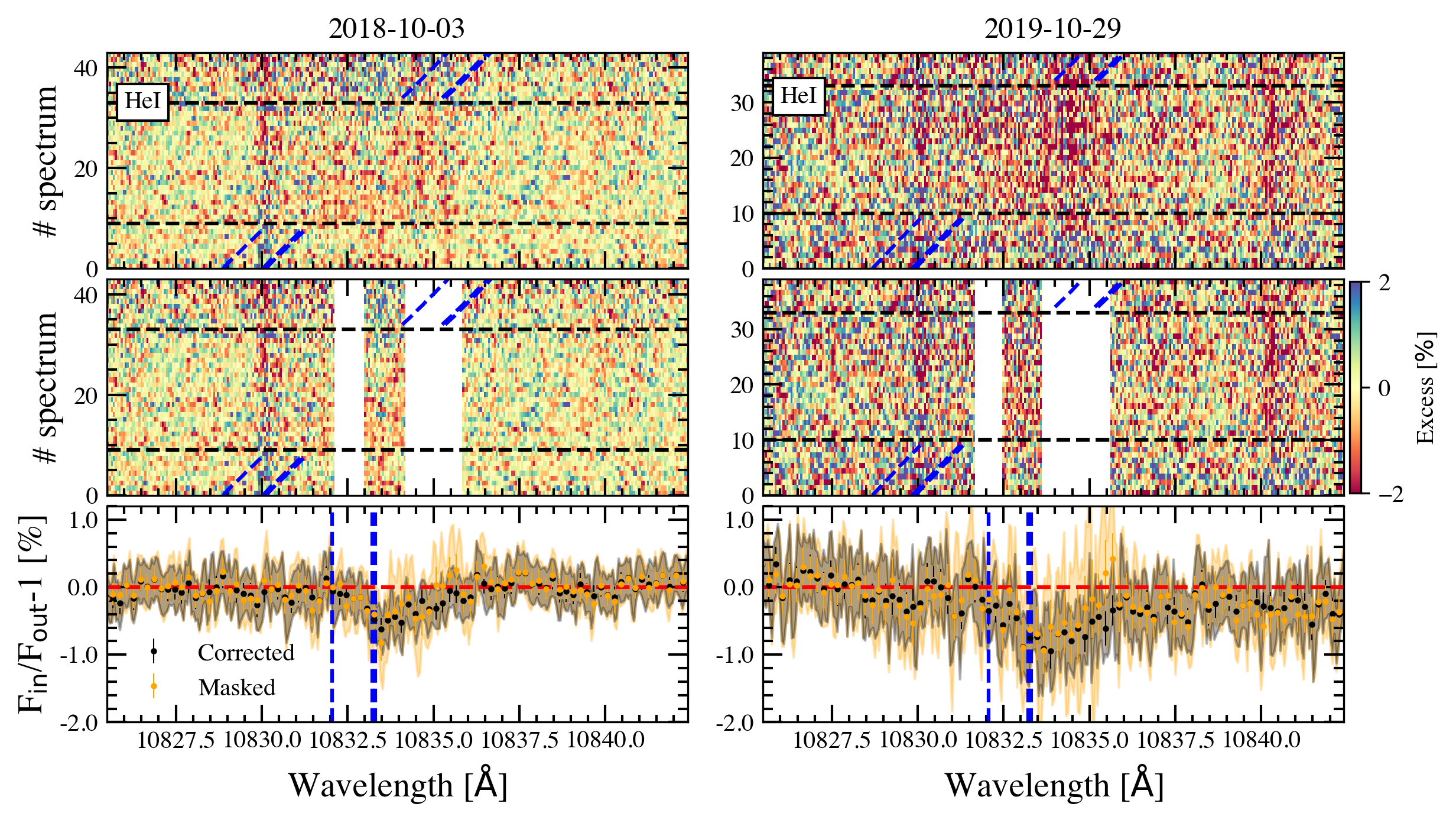}
\caption{{Results around the \ion{He}{i} lines for the first ({\em left panel}) and the second night ({\em right panel}). 
{\em Top row}: tomography map obtained correcting the telluric absorption and emission lines. Middle row: same as top row but masking the telluric contamination (white regions). 
{\em Bottom row}: transmission spectrum obtained correcting the telluric lines (black), and masking the lines (orange). The light coloured region show the error bars of the original data sampling and the data binned by $0.2$\,{\AA} are shown in dots.}}
\label{fig:ts_HeI}
\end{figure*}

\begin{figure}[]
\centering
\includegraphics[width=0.5\textwidth]{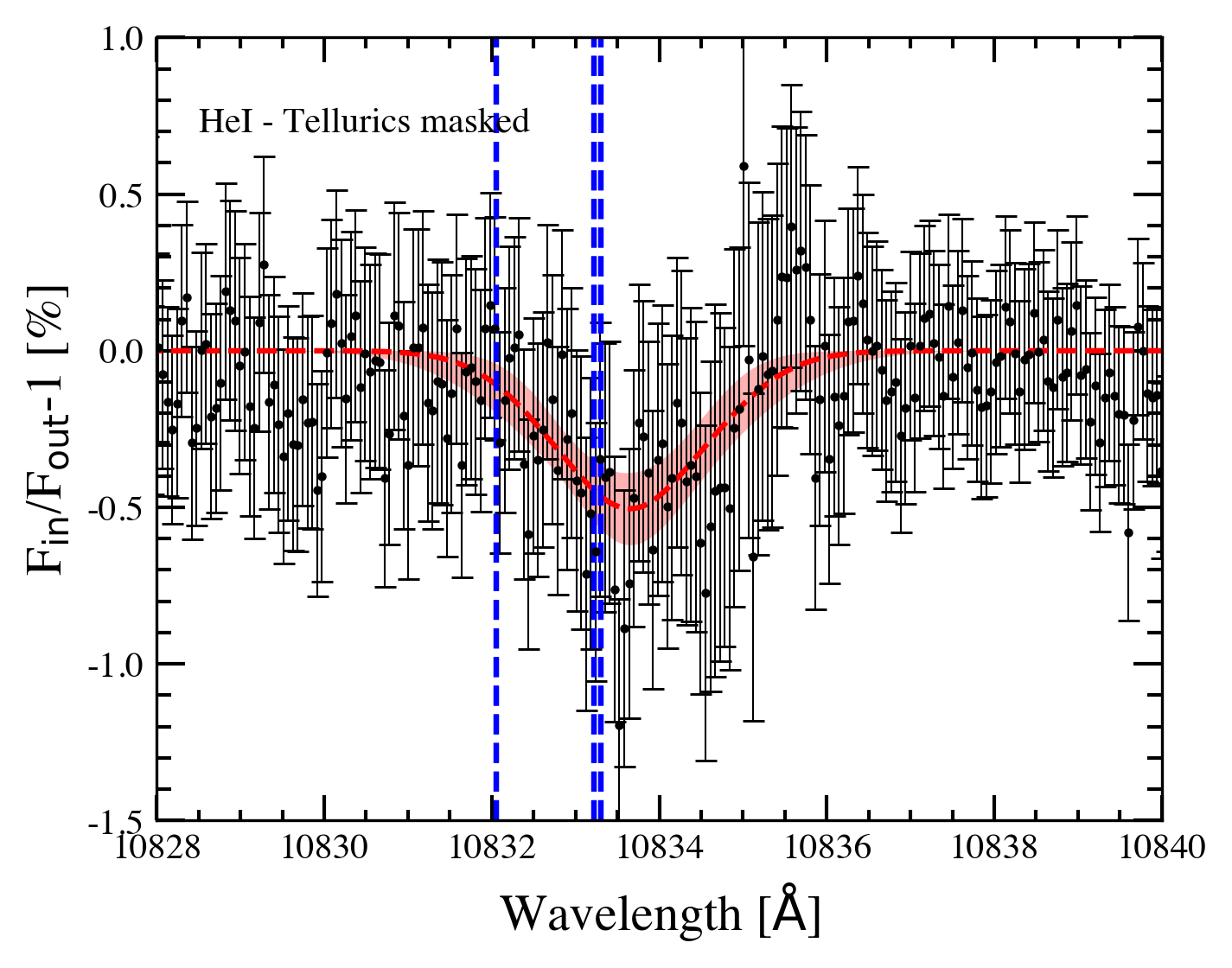}
\caption{CARMENES transmission spectrum of WASP-76b around the \ion{He}{i} lines after combining the results of the two nights shown in Fig.~\ref{fig:ts_HeI}, obtained by masking the telluric contamination. The red dashed line shows the best-fit Gaussian profile, and the light red regions indicate 1$\sigma$ uncertainties. The blue dashed vertical lines show the expected positions of the \ion{He}{i} lines.}
\label{fig:ts_HeI_comb}
\end{figure}

The individual results obtained in the analysis of the CARMENES observations of WASP-76b around the near-infrared \ion{He}{i} lines are shown in Fig.~\ref{fig:ts_HeI}, and the combined transmission spectrum is presented in Fig.~\ref{fig:ts_HeI_comb}. Both nights show a broad and shallow ($<1~\%$) red-shifted absorption feature at the \ion{He}{i} line positions. As telluric residuals remain in the tomography maps after correcting the emission and absorption from the Earth's atmosphere (top row of Fig.~\ref{fig:ts_HeI}), we additionally extracted the transmission spectrum by masking those regions. We observed that, although the latter result was noisier, both transmission spectra were consistent. We measured the \ion{He}{i} absorption feature observed in the transmission spectrum by fitting a Gaussian profile as detailed in Sect.~\ref{sec:results_Ca}. 
For the case 
with masked telluric contamination, 
we found an excess of absorption of $0.52\pm0.12\,\%$ ($\sim4\sigma$), FWHM of $2.1\pm0.6$\,{\AA}, and a redshift of $10\pm5$\,km\,s$^{-1}$. For the case 
with corrected contamination, 
the transmission spectrum shows consistent absorption excess ($0.55\pm0.06\,\%$) and width ($2.9\pm0.4$\,{\AA}), but it is more highly redshifted ($\sim18$\,km\,s$^{-1}$) due to the strong telluric residuals that remain in that position. The best fit Gaussian profile for the case where the telluric contamination was masked is shown in Fig.~\ref{fig:ts_HeI_comb}, and the probability distributions in Fig.~\ref{fig:corner}. Due to the remaining telluric residuals, as well as the small number of pixels that represent the transmission spectrum when masking the contamination, we conservatively used our results to derive an upper limit of $0.88\,\%$ for the \ion{He}{i} absorption in the atmosphere of WASP-76b at the 3$\sigma$ confidence level, derived from the MCMC probability distribution of the Gaussian profile fitting procedure. If the upper limit is calculated as the 1-pixel dispersion in the transmission spectrum, we obtain a smaller value of $0.67\,\%$, but 
the most conservative value ($<0.88\,\%$) is preferred for this particular case.

In addition to the surrounding telluric contamination that affects the transmission spectrum around \ion{He}{i}, we explored the possibility that the stellar companion at $0.44$\,arcsec \citep{Southworth2020} could slightly contaminate the transmission spectrum of WASP-76b. We observed that on the second night the absorption was noticeably deeper than on the first night. One possibility is that the spectrum of WASP-76 is contaminated by the stellar companion, 
which affects the two transits differently due to differences in activity level. 
However, the \ion{He}{i} lines are less sensitive to stellar activity than other lines \citep{Cauley2018}. We compute the transmission spectrum assuming different $K_{\rm p}$ values and look for the one showing the strongest absorption. The feature appears in a wide range of $K_{\rm p}$, from --100 to +300\,km\,s$^{-1}$ (see Fig.~\ref{fig:Kp_HeI}), which supports the notion that the origin of the \ion{He}{i} absorption feature could be due to telluric residuals, contamination from the stellar companion, or both. Further observations are needed to confirm the origin of our findings.

We explored future opportunities to confirm the presence of \ion{He}{i} in the atmosphere of WASP-76b using transit observations with CARMENES. The \ion{He}{i} lines are located between telluric OH emission lines (see Fig.~\ref{fig:HeI_star}) and, therefore, the system velocity ($\gamma$) and the barycentric velocity play a big role in the way in which one may find optimal transits to perform these studies. Consequently, only a few opportunities are accessible during the year. In terms of the telluric emission lines position, the second night analysed here represents one of the most optimal situations to study the \ion{He}{i} lines in WASP-76b. Unfortunately, due to the weather conditions, these observations had low S/N. The next three optimal transit opportunities will occur during the nights of 2 October 2021, 31 October 2021, and 28 October 2021. The combination of one additional transit observation obtained under good weather conditions would significantly increase the robustness of the results, decreasing the impact of the telluric residuals.

\begin{figure}[]
\centering
\includegraphics[width=0.5\textwidth]{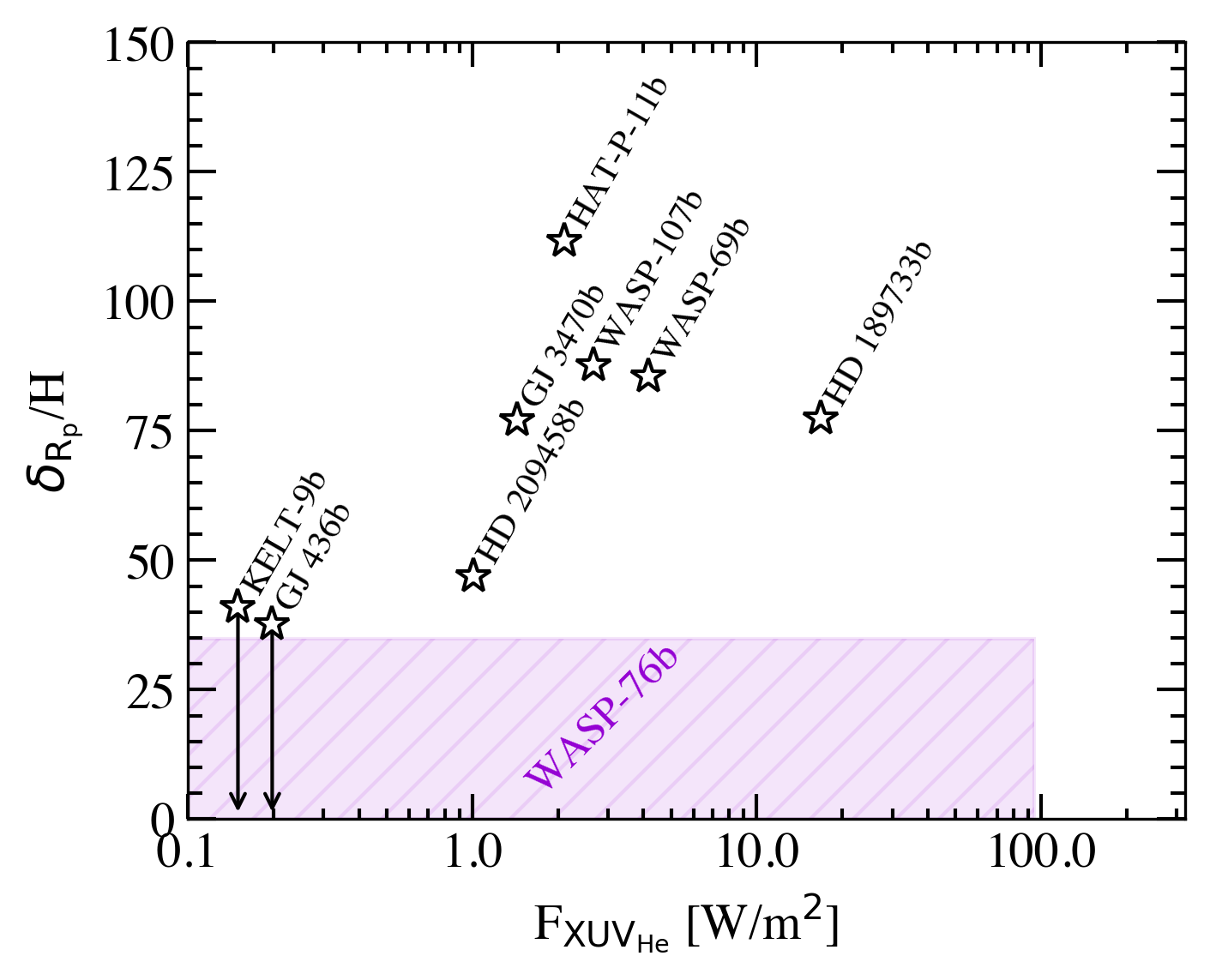}
\caption{\ion{He}{i} transmission signals measured using observations at high spectral resolution with CARMENES, as a function of the XUV$_{\rm He}$ flux ($\lambda <504$\,{\AA}) received by the planet. The vertical axis shows the equivalent height of the \ion{He}{i} atmosphere $\delta_{R_{\rm p}}$, normalised by its atmospheric scale height in the lower atmosphere, $H$, calculated assuming $\mu=2.3$ and the equilibrium temperature of the planets. The coloured region shows the constrained ranges defined by the \ion{He}{i} absorption and XUV$_{\rm He}$ flux upper limits measured for WASP-76b. The information on the other exoplanets is extracted from \citet{Nortmann2018Science}, \citet{AlonsoFloriano2019HD209}, and \citet{Palle2020a}. 
The black arrows show upper limits on the measured \ion{He}{i} absorption of KELT-9\,b and GJ~436\,b.}
\label{fig:XUV}
\end{figure}

Following the methodology presented by \citet{Nortmann2018Science}, we calculated the XUV flux that WASP-76b receives on its atmosphere. Due to the distance to the host star (195\,pc), we did not reliable detect WASP-76 in
EPIC/{\em XMM-Newton} data. The resulting values can, therefore, only be considered as upper limits. The exoplanet receives $F_{\rm X}< 18$\,W\,m$^{-2}$ and $F_{\rm EUV_{He}}<76$\,W\,m$^{-2}$, with $F_{\rm X}$ calculated considering the photons in the wavelength range 5--100\,{\AA}, and $F_{\rm EUV_{He}}$ considering those in 100--504\,{\AA}. 
These fluxes translate into a $F_{\rm XUV_{He}}<94$\,W\,m$^{-2}$, including only the photons required to ionise \ion{He}{i}. With a total flux in the 5--920\,{\AA} range of $F_{\rm XUV} < 148$~W\,m$^{-2}$, the energy-limited mass loss rate (\citealt{SanzForcada2011} and references therein) is $< 150\,{\rm M}_{\oplus}$\,Gyr$^{-1}$ (or $2.8\times10^{13}$\,g\,s$^{-1}$). Although other late F stars with known X-ray spectra exist (e.g. \object{$\tau$~Boo} and \object{$\iota$~Hor}), they can not be used as a proxy to estimate the X-ray flux of WASP-76 due to their different activity levels, as inferred from $L_{\rm X}/L_{\rm bol}$. These objects are more active than WASP-76, and finding late F stars with $\log L_{\rm X}/L_{\rm bol} < -6$, as it is the case of WASP-76, is a challenging task. For this reason, our low-significance measurement is preferred, and it is expected to not differ much from the actual level of X-ray emission of WASP-76. When compared with the \ion{He}{i} absorption measurements on the atmospheres of other planets, the absorption and XUV upper limits estimated here show that WASP-76b could possess a \ion{He}{i} envelope with possible material escaping from its exosphere (see Fig.~\ref{fig:XUV}).

\subsection{Bootstrap analysis of significance}

To check the origin of the measured absorption features in the transmission spectrum, we additionally applied the empirical Monte Carlo (EMC) or bootstrap analysis, which is based on measuring the absorption in the transmission spectrum when computed assuming thousands of different in- and out-of-transit samples \citep{2008Redfield}. 
If the absorption observed in the transmission spectrum has planetary origin, 
it would be reproduced 
only when the correct in-transit and out-of-transit samples are used. 
We considered three different scenarios: `in-in', `out-out', and `in-out'. In the first two scenarios, both virtual in- and out-of-transit samples contain real in-transit and out-of-transit spectra only, respectively, which were randomised. 
In these two cases we expected to find most of the measurements centred at null absorption. The `in-out' scenario corresponds to the absorption case, where the samples were still randomised but contained the correct in-transit and out-of-transit spectra. 
In this case, if there were any absorption from the exoplanet atmosphere, we would find the distribution of the measurements centred at a negative value significantly displaced from zero. For these calculations we used 20\,000 realisations. 

\begin{figure}[]
\centering
\includegraphics[width=0.4\textwidth]{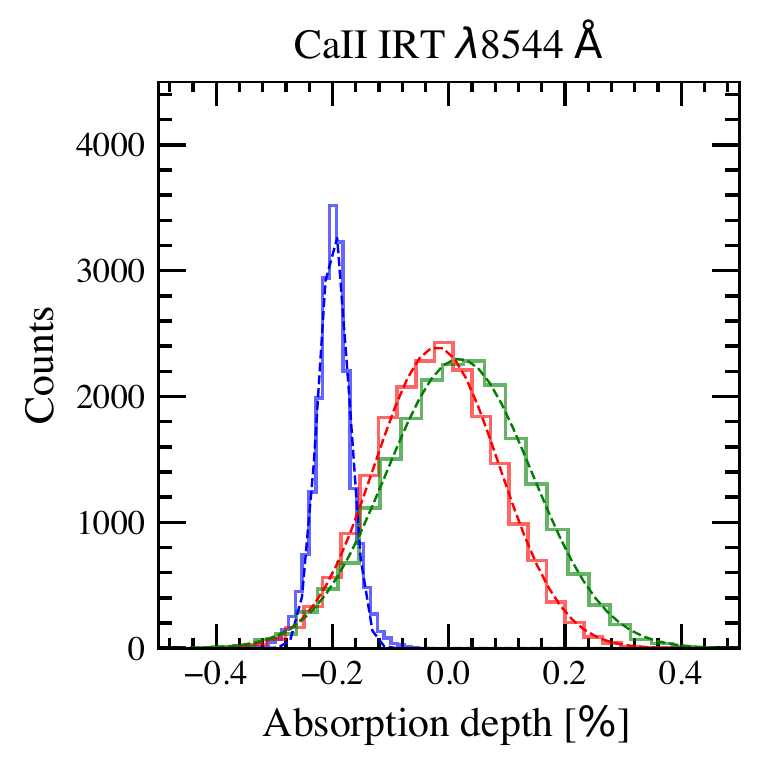}
\caption{EMC distributions around the \ion{Ca}{ii} IRT $\lambda$8544\,{\AA} line for both nights combined. The results of the `in-in' sample are shown in red, the `out-out' in green, and the `in-out' in blue. The dashed lines show the best-fit Gaussian profiles of the distributions.}
\label{fig:emc_Ca2}
\end{figure}

For the \ion{Ca}{ii} IRT lines, we applied the EMC analysis using a $2$\,{\AA} bandwidth. As expected, in all cases, we observed that the `in-in' and `out-out' distributions are centred at 0\,\% absorption, while the `in-out' distributions of the different lines are shifted to negative values with respect to the reference samples, suggesting excess of absorption. Following \citet{Wytt2015} and \citet{Seidel2019}, we inferred the significance of this shift by considering the error of the measurement as the standard deviation of the `out-out' distribution, and scaling it to the number of spectra. The absorption measured with the EMC is significant ($3\sigma$) only for the strongest line of the triplet (see Fig.~\ref{fig:emc_Ca2}, as observed in the analysis of the individual lines in the transmission spectrum. This analysis discarded a spurious origin of this \ion{Ca}{ii} IRT line. The distributions of of the other \ion{Ca}{ii} IRT lines are presented in Fig.~\ref{fig:emc1}.

We also applied this analysis to the \ion{He}{i} lines and found that, with a $5$\,{\AA} bandwidth, the absorption signal was inferred for the `in-out' sample (see distribution in Fig.~\ref{fig:emc1}) at $\sim4\sigma$. However, the `out-out' distribution was also slightly shifted from the 0\,\% absorption, which suggested the presence of systematic noise. As discussed in Sect.~\ref{sec:results_HeI}, additional observations are needed to robustly confirm the origin of this absorption.

\subsection{Revisiting H$\alpha$, \ion{Na}{i}, \ion{Li}{i}, and \ion{K}{i}}

Atmospheric studies at high spectral resolution have shown that WASP-76b possess a transmission spectrum rich in metal lines \citep{Seidel2019,Zak2019,Ehrenreich2020,Tabernero2020}, which includes Na, K, Fe, and Li, among several other species. On the other hand, and in contrast with other ultra-hot Jupiters, its transmission spectrum does not show detectable H in the Balmer lines. 

The transmission spectrum of WASP-76b obtained with two transit observations with CARMENES shows no detectable H$\alpha$, \ion{Li}{i}, or \ion{K}{i} absorption due to the low S/N of the observations (see Fig.~\ref{fig:revis_res}). For this same reason, around the \ion{Na}{i} doublet only the \ion{Na}{i} D$_1$ line is detected tentatively (at $2\sigma$). For this latter case, the transmission spectrum was extracted using three different approaches: 
($i$) considering all data, 
($ii$) attempting the correction of the telluric \ion{Na}{i} emission, and ($iii$) masking the strong contamination in the line cores. All approaches give consistent transmission spectra (see Fig.~\ref{fig:ts_NaI}), which show too low S/N to firmly detect the \ion{Na}{i} absorption reported in previous studies.

Based on the results obtained by \citet{Tabernero2020} with ESPRESSO observations, the contrast of the \ion{Li}{i} and \ion{K}{i} absorption lines detected in WASP-76b is around $0.2\,\%$. The transmission spectrum obtained here shows uncertainties of the same order of magnitude in the regions surrounding these lines and, therefore, the absorption can not be distinguished from the noise. We were only able to place an upper limit of $0.39\,\%$ and $0.43\,\%$ (at $3\sigma$ confidence level) on these lines, respectively, which are consistent with previous findings. These upper limits were obtained from the null hypothesis, using the 1-pixel dispersion calculated in the surroundings of the line's position in the transmission spectrum. The H$\alpha$ absorption was not detected by \citet{Zak2019} nor \citet{Essen2020}, but it was detected only at $4\sigma$ on one night by \citet{Tabernero2020}. Our observations, although at lower S/N, are also consistent with a non-detection of H$\alpha$ in the atmosphere of WASP-76b with an upper limit of $0.47\,\%$. 
Finally, \citet{Tabernero2020} detected the \ion{Na}{i} doublet with depths of $0.449\pm0.049\,\%$ and $0.246\pm0.037\,\%$ (D$_2$) and $0.385\pm0.051\,\%$ and $0.294\pm0.042\,\%$ (D$_1$) on their two nights of observation, respectively. 
\citet{Seidel2019}, using HARPS, measured a depth of $0.373\pm0.091\,\%$ (D$_2$) and $0.508\pm0.083\,\%$ (D$_1$). These results were revised by \citet{Seidel2021}, after the discovery of the stellar companion WASP-76B and the subsequent improvement of the system parameters, reporting changes $<1\%$ compared to the spectrum published by \citet{Seidel2019}. After masking the stellar line cores, the transmission spectrum around the \ion{Na}{i} doublet of our CARMENES observations only shows a tentative absorption around the \ion{Na}{i} D$_1$ line of $0.2\pm0.1\,\%$, consistent with \citet{Tabernero2020} within $\sim 1\sigma$, but differing by $2.4~\sigma$ from the D$_1$ line measurements reported by \citet{Seidel2019}. The $3\sigma$ upper limit measured for the \ion{Na}{i} doublet is $0.48\,\%$, calculated from the 1-pixel dispersion, and $0.66\,\%$ and $0.58\,\%$ for the D$_2$ and D$_1$ lines, respectively, using the $3\sigma$ uncertainties from the MCMC Gaussian fitting. These values are consistent with the measurements reported in previous studies. The upper limits are summarised in Table~\ref{tab:upper limits}.

\begin{table}[]
\centering
\caption{Summary of the $3\sigma$ upper limits reported for the different species.}
\begin{tabular}{lc}
\hline
\hline
\noalign{\smallskip}
Species & $3\sigma$ upper limit  [\%]\\
\noalign{\smallskip}
\hline 
\noalign{\smallskip}

\ion{He}{i}  &  0.67/0.88$^a$\\

\ion{Li}{i} & 0.39 \\

\ion{K}{i} & 0.43 \\

H$\alpha$ & 0.47 \\

\ion{Na}{i} D$_2$ & 0.48/0.66$^a$ \\

\ion{Na}{i} D$_1$ & 0.48/0.58$^a$\\

\noalign{\smallskip}
\hline
\end{tabular}\\
\label{tab:upper limits}
\tablefoot{\tablefoottext{a} Obtained from the MCMC Gaussian fit posteriors, assuming the $3\sigma$ upper limit of the absorption excess. The rest of the values are obtained assuming the null hypothesis with the 1-pixel dispersion.}
\end{table}

\section{Conclusions} \label{sec:conclusions}

We present the transmission spectrum of WASP-76b obtained with two transit observations with CARMENES. Here, we focus on the search for Ca$^+$ and He by taking advantage of the red-optical and near-infrared coverage of CARMENES. In particular, we analyse for the first time in this planet the \ion{Ca}{ii} IRT at 8500\,{\AA} and the \ion{He}{i} triplet at 10830\,{\AA}. 

The analysis of the individual \ion{Ca}{ii} IRT lines shows a detection ($5.6\sigma$) of the strongest line of the triplet at $8544$\,{\AA}, but the two fainter lines are not detected with high significance ($\lesssim 3\sigma$). Using the cross-correlation technique, we probe the combined signal of the three lines, resulting in a $7\sigma$ detection of \ion{Ca}{ii} absorption. After MASCARA-2b \citep{Casasayas2019}, KELT-9b and WASP-33b \citep{Yan2019}, WASP-76b is the coolest highly irradiated exoplanet with a detection of \ion{Ca}{ii} IRT absorption. The cross-correlation is applied using several atmospheric models computed under different assumptions for rotational broadening and temperatures. The results show very similar significance (6--8$\sigma$) and shape. The transmission spectrum around the strongest \ion{Ca}{ii} IRT line at $8544$\,{\AA} suggests that broadening is needed to explain the line profile, which is consistent with the results obtained by \citet{Seidel2019} around the \ion{Na}{i} lines. We consider the line broadening produced by the tidally locked rotation of the planet, but still additional broadening is observed in the lines, which could be attributable to vertical winds or atmospheric turbulence. Moreover, the results suggest temperatures higher than 2200\,K (the equilibrium temperature of the planet), probably because these lines are created in a hotter region of the atmosphere, thus pointing to a possible temperature inversion in this atmosphere. Our detection of Ca$^+$ in the atmosphere of WASP-76b is consistent with the results presented by \citet{Tabernero2020}, who detected the same species, but in the \ion{Ca}{ii} H\&K lines. 

We also use the NIR channel of CARMENES to analyse the \ion{He}{i} triplet lines. Both nights showed telluric contamination close to the expected position of the exoplanet absorption. Following two different approaches, the transmission spectra of the two nights show a broad red-shifted ($\sim10$\,km\,s$^{-1}$) absorption feature of $\sim0.5\,\%$ depth. However, the modest S/N and the telluric contamination in the observations prevent us from robustly detecting the presence of He in the atmosphere of WASP-76b. Nevertheless, we are able to provide an upper limit of $0.9\,\%$ on the absorption depth. Interestingly, this upper limit points to the possibility of this planet having an evaporating He-rich envelope that, if confirmed, would be the first case in an ultra-hot Jupiter. Therefore, this planet is a candidate for further studies to confirm and improve our findings.

Finally, we explore other regions of the transmission spectrum, in particular, the H$\alpha$, \ion{Na}{i}, \ion{Li}{i}, and \ion{K}{I} lines. H$\alpha$ was studied by \citet{Zak2019} and \citet{Tabernero2020}, who reported a non-detection and an inconclusive result, respectively. \ion{Li}{i} and \ion{K}{i} absorption were detected by \citet{Tabernero2020}, while 
they along with \citet{Zak2019} and \citet{Seidel2019} found Na in the atmosphere of WASP-76b. With our observations, we are not able to detect any of these species. Comparing the absorption level of previous findings with the transmission spectra obtained here, we conclude that the S/N of our CARMENES spectra is not sufficient to distinguish these species from the noise level. However, the upper limits are consistent with the reported detections. In the case of \ion{Na}{i}, low S/N and telluric contamination is observed in the stellar line cores, preventing the atmospheric absorption from being recovered. 

The transmission spectrum of WASP-76b is characterised by strong absorption features due to the combined effects of a large atmospheric scale height and a rather clear sky. In fact, the high atmospheric temperatures inhibit the precipitation of volatiles and the formation of clouds \citep{Wakeford2017}, although refractory clouds are still possible \citep{Sing2013}. In particular, the inventories of chemical species that have been reported for WASP-76b and WASP-121b are remarkably similar. 
The well investigated WASP-121b planet is one of the closest neighbours to WASP-76b in the equilibrium temperature versus mean density plot (Fig.~\ref{fig:UHJ_context}), and they both orbit late F-type stars with slightly super-solar metallicities \citep{Tabernero2020}. The similarities in their atmospheric composition could point towards similar formation pathways \citep{Turrini2015}.

\begin{acknowledgements}

We thank the referee for the helpful report, and constructive remarks on this manuscript. 

CARMENES is an instrument at the Centro Astron\'omico Hispano-Alem\'an (CAHA) at Calar Alto (Almer\'{\i}a, Spain), operated jointly by the Junta de Andaluc\'ia and the Instituto de Astrof\'isica de Andaluc\'ia (CSIC).
  
  CARMENES was funded by the Max-Planck-Gesellschaft (MPG), 
  the Consejo Superior de Investigaciones Cient\'{\i}ficas (CSIC),
  the Ministerio de Econom\'ia y Competitividad (MINECO) and the European Regional Development Fund (ERDF) through projects FICTS-2011-02, ICTS-2017-07-CAHA-4, and CAHA16-CE-3978, 
  and the members of the CARMENES Consortium 
  (Max-Planck-Institut f\"ur Astronomie,
  Instituto de Astrof\'{\i}sica de Andaluc\'{\i}a,
  Landessternwarte K\"onigstuhl,
  Institut de Ci\`encies de l'Espai,
  Institut f\"ur Astrophysik G\"ottingen,
  Universidad Complutense de Madrid,
  Th\"uringer Landessternwarte Tautenburg,
  Instituto de Astrof\'{\i}sica de Canarias,
  Hamburger Sternwarte,
  Centro de Astrobiolog\'{\i}a and
  Centro Astron\'omico Hispano-Alem\'an), 
  with additional contributions by the MINECO, 
  the Deutsche Forschungsgemeinschaft through the Major Research Instrumentation Programme and Research Unit FOR2544 ``Blue Planets around Red Stars'', 
  the Klaus Tschira Stiftung, 
  the states of Baden-W\"urttemberg and Niedersachsen, 
  and by the Junta de Andaluc\'{\i}a.
  
  This work was based on data from the CARMENES data archive at CAB (CSIC-INTA).

We acknowledge funding from the European Research Council under the European Union's Horizon 2020 research and innovation program under grant agreement No.~694513, 
the Agencia Estatal de Investigaci\'on of the Ministerio de Ciencia, Innovaci\'on y Universidades and the ERDF through projects 
  PID2019-109522GB-C5[1:4]/AEI/10.13039/501100011033,	
  PID2019-110689RB-I00/AEI/10.13039/501100011033,
  ESP2017-87143-R, and ESP2016-80435-C2-2-R,
and the Centre of Excellence ``Severo Ochoa'' and ``Mar\'ia de Maeztu'' awards to the Instituto de Astrof\'isica de Canarias (CEX2019-000920-S), Instituto de Astrof\'isica de Andaluc\'ia (SEV-2017-0709), and Centro de Astrobiolog\'ia (MDM-2017-0737), and the Generalitat de Catalunya/CERCA programme. T.H. acknowledges support by the European Research Council under the Horizon 2020 Framework Program via the ERC Advanced Grant Origins 83 24 28. G. M. has received funding from the European Union's Horizon 2020 research and innovation programme under the Marie Skłodowska-Curie grant agreement No. 895525.
  
This work made use of PyAstronomy \citep{PyAstronomy2019ascl.soft06010C} and of the VALD database, operated at Uppsala University, the Institute of Astronomy RAS in Moscow, and the University of Vienna.

\end{acknowledgements}

\bibliographystyle{aa} 
\bibliography{biblio} 

\onecolumn
\begin{appendix} 

\section{Additional figures}
\label{app:addfig}

\subsection{Transmission spectra}

\begin{figure*}[h]
\centering
\includegraphics[width=1\textwidth]{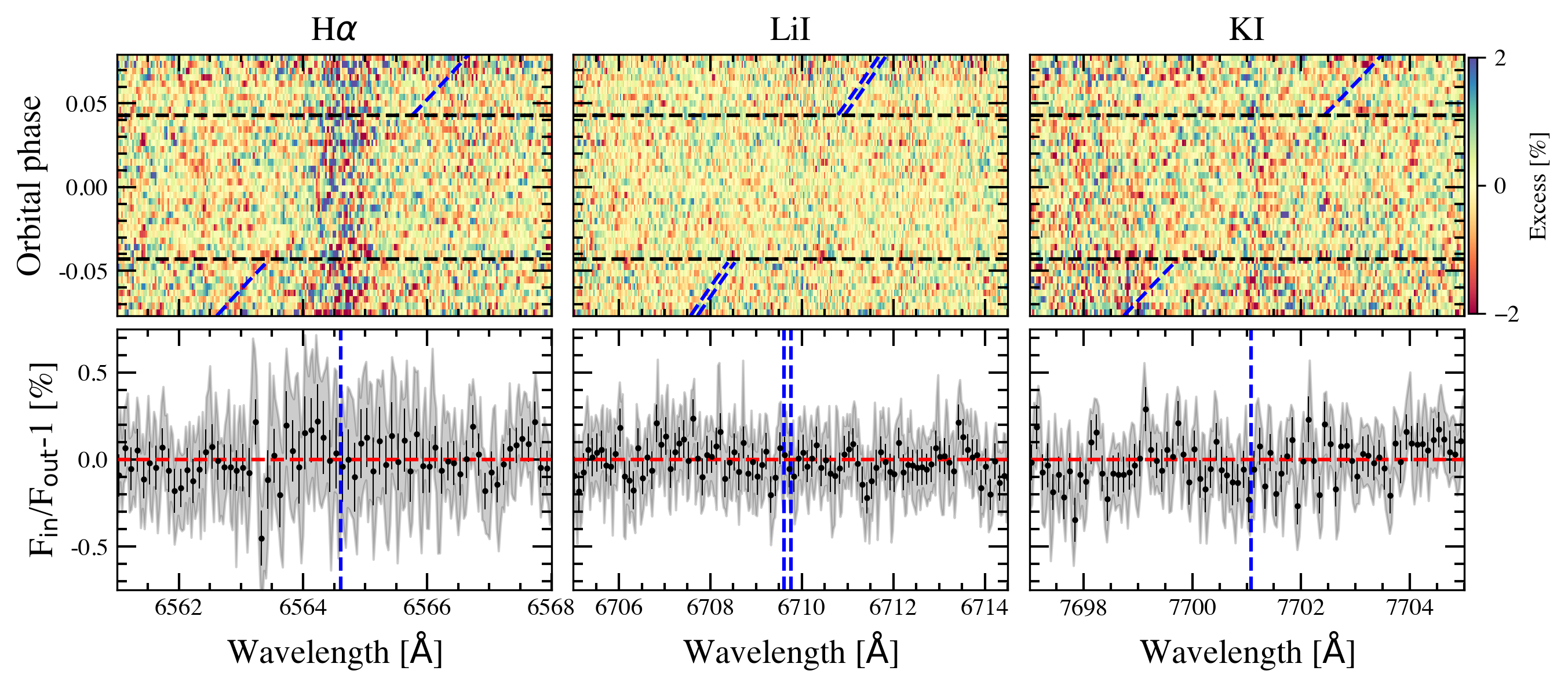}
\caption{Same as Fig.~\ref{fig:TS_CaIRT}, but for H$\alpha$, \ion{Li}{i}, and \ion{K}{i}.}
\label{fig:revis_res}
\end{figure*}

\begin{figure}[h]
\centering
\includegraphics[width=0.5\textwidth]{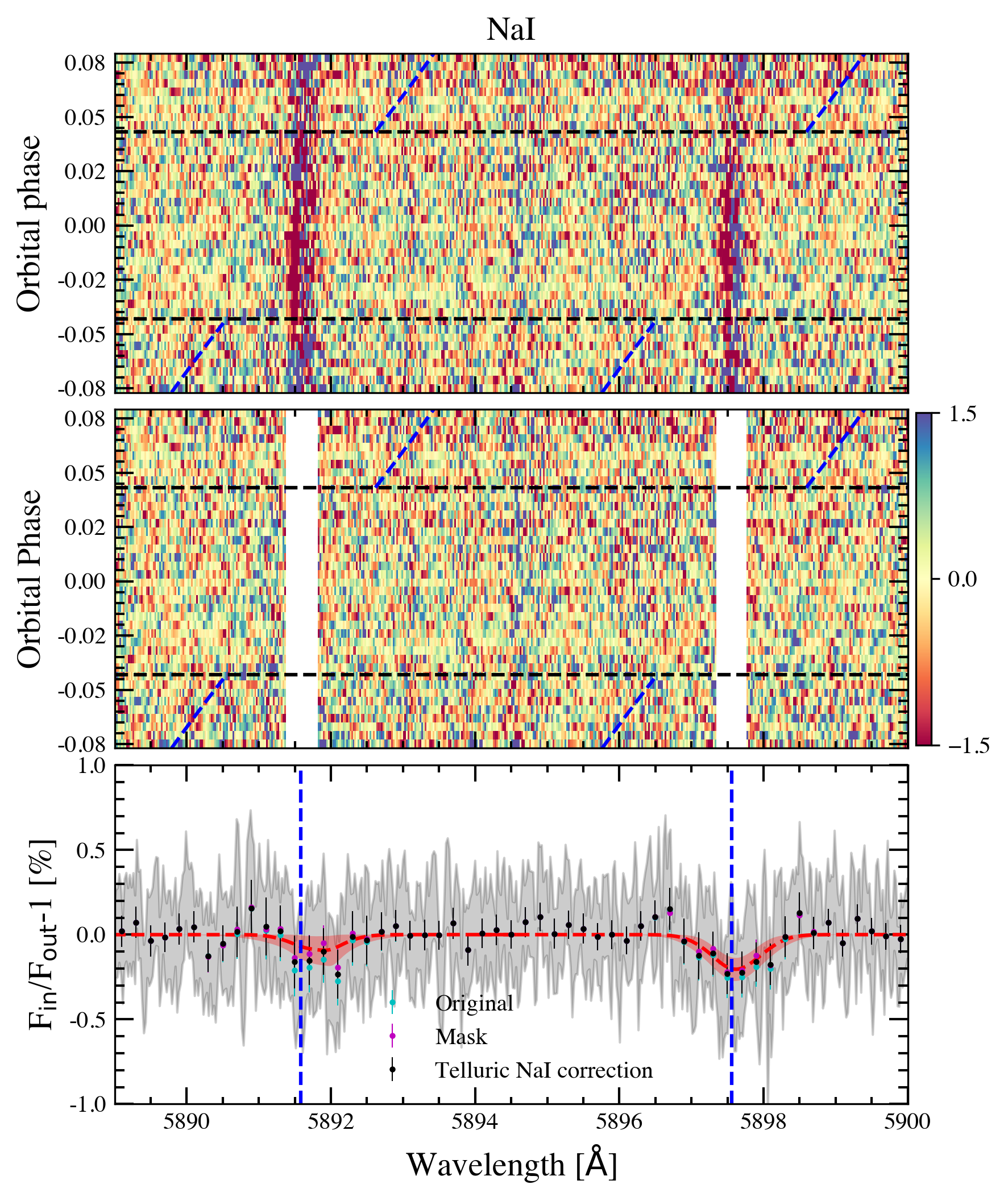}
\caption{Same as Fig.~\ref{fig:TS_CaIRT}, but for the \ion{Na}{i} D$_1$ and D$_2$ lines. The middle panel shows the same as the top panel but after masking the contaminated stellar line cores around $\pm10$\,km\,s$^{-1}$ (white regions). The bottom panel shows the transmission spectrum obtained using different methodologies: using all data from the top panel (cyan), masking the centre of the stellar \ion{Na}{i} lines core as in the middle panel (magenta), and correcting the telluric \ion{Na}{i} emission contamination (black). The transmission spectrum is binned by $0.2$\,{\AA} for better visualisation. The best-fit Gaussian profile is shown in red.}
\label{fig:ts_NaI}
\end{figure}

\clearpage

\subsection{Cross-correlation results}

\begin{figure*}[h]
\centering
\includegraphics[width=0.98\textwidth]{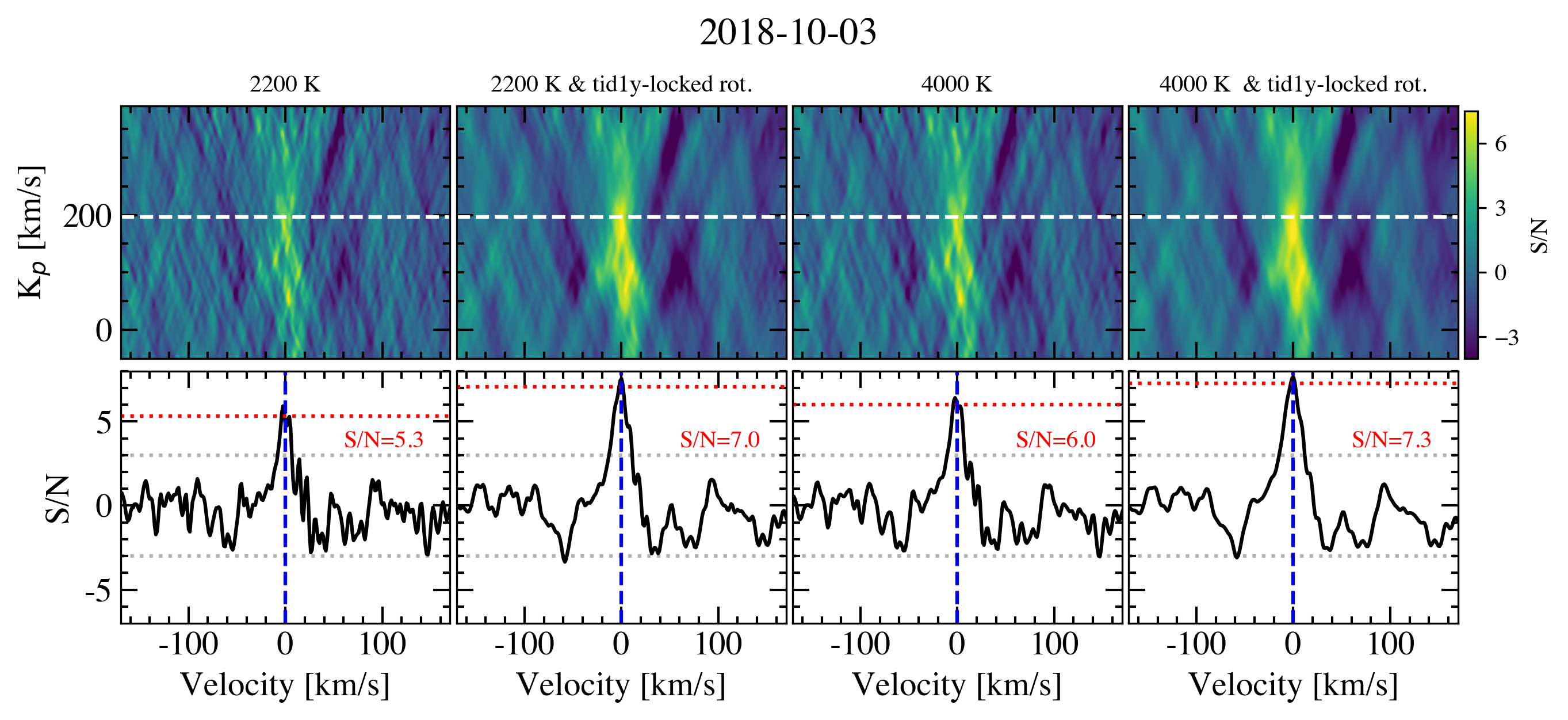}
\includegraphics[width=0.98\textwidth]{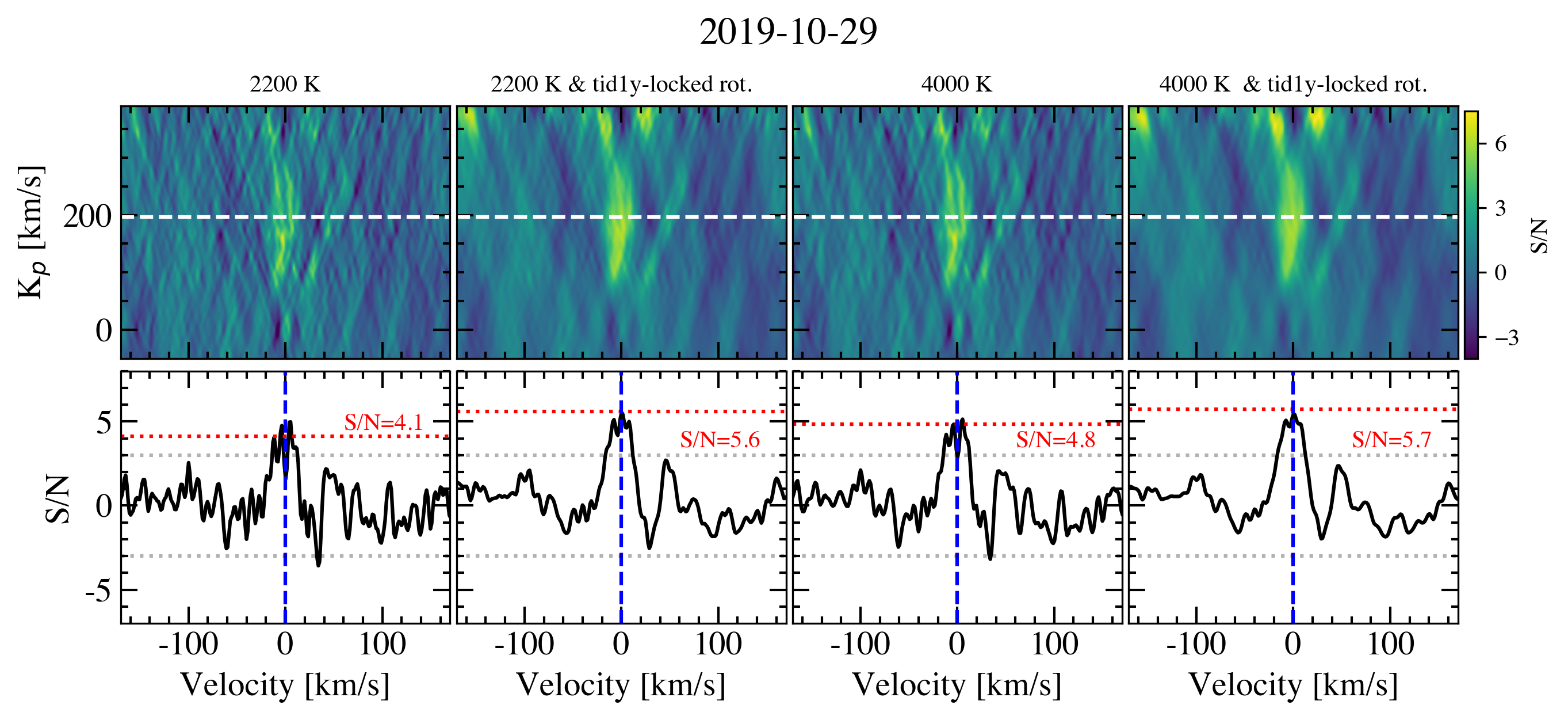}
\caption{Same as Fig.~\ref{fig:CC_CaIRT}, but for the two individual nights.}
\label{fig:cc_Ca_ind}
\end{figure*}

\clearpage

\subsection{\ion{He}{i} tests}

\begin{figure*}[h]
\centering
\includegraphics[width=0.98\textwidth]{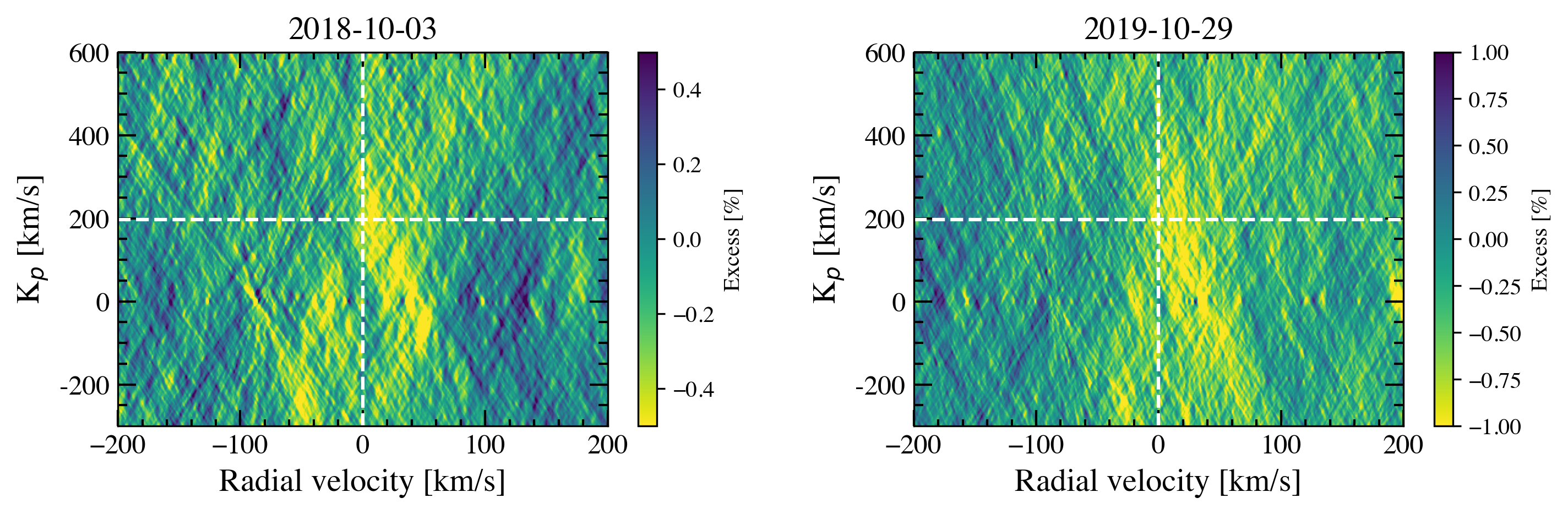}
\caption{$K_p$-velocity map around the \ion{He}{i} lines for each individual night, obtained applying the telluric correction. The horizontal and vertical white-dashed lines show the expected velocity position of the signal ($0$\,km\,s$^{-1}$) and the expected $K_p$ value ($196.52$\,km\,s$^{-1}$), respectively.}
\label{fig:Kp_HeI}
\end{figure*}

\clearpage

\subsection{MCMC probability distributions}

\begin{figure*}[h]
\centering
\includegraphics[width=0.45\textwidth]{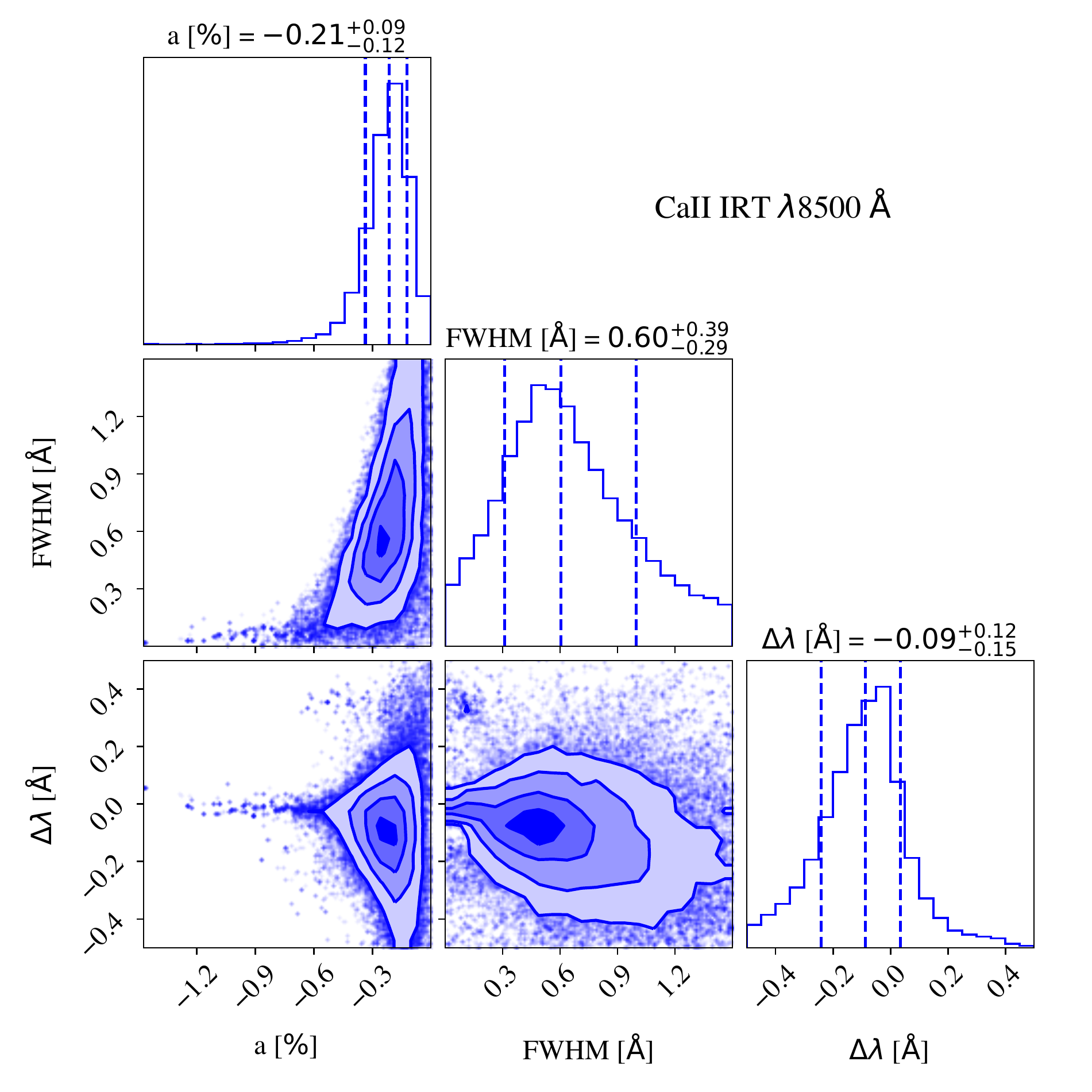}
\includegraphics[width=0.45\textwidth]{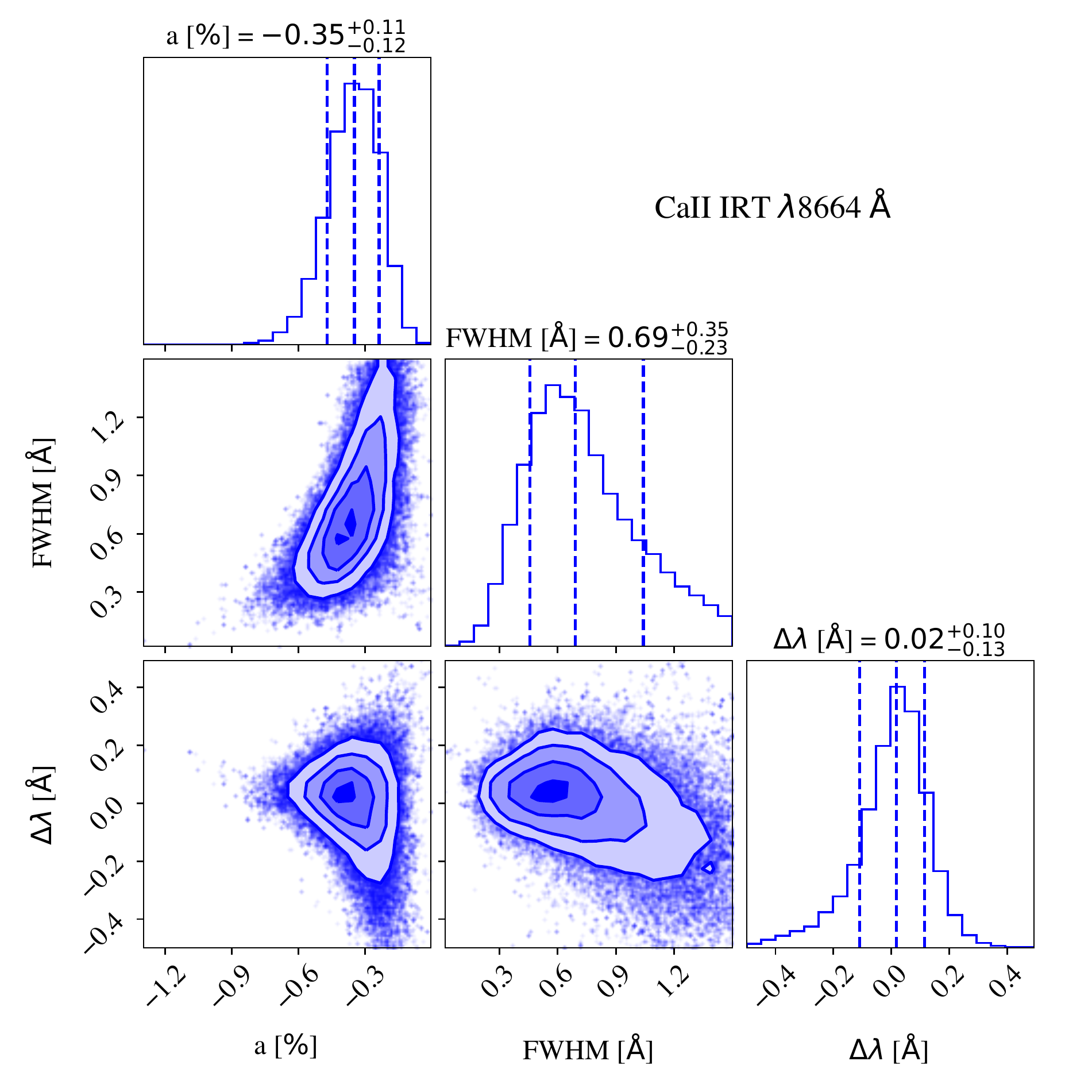}
\includegraphics[width=0.45\textwidth]{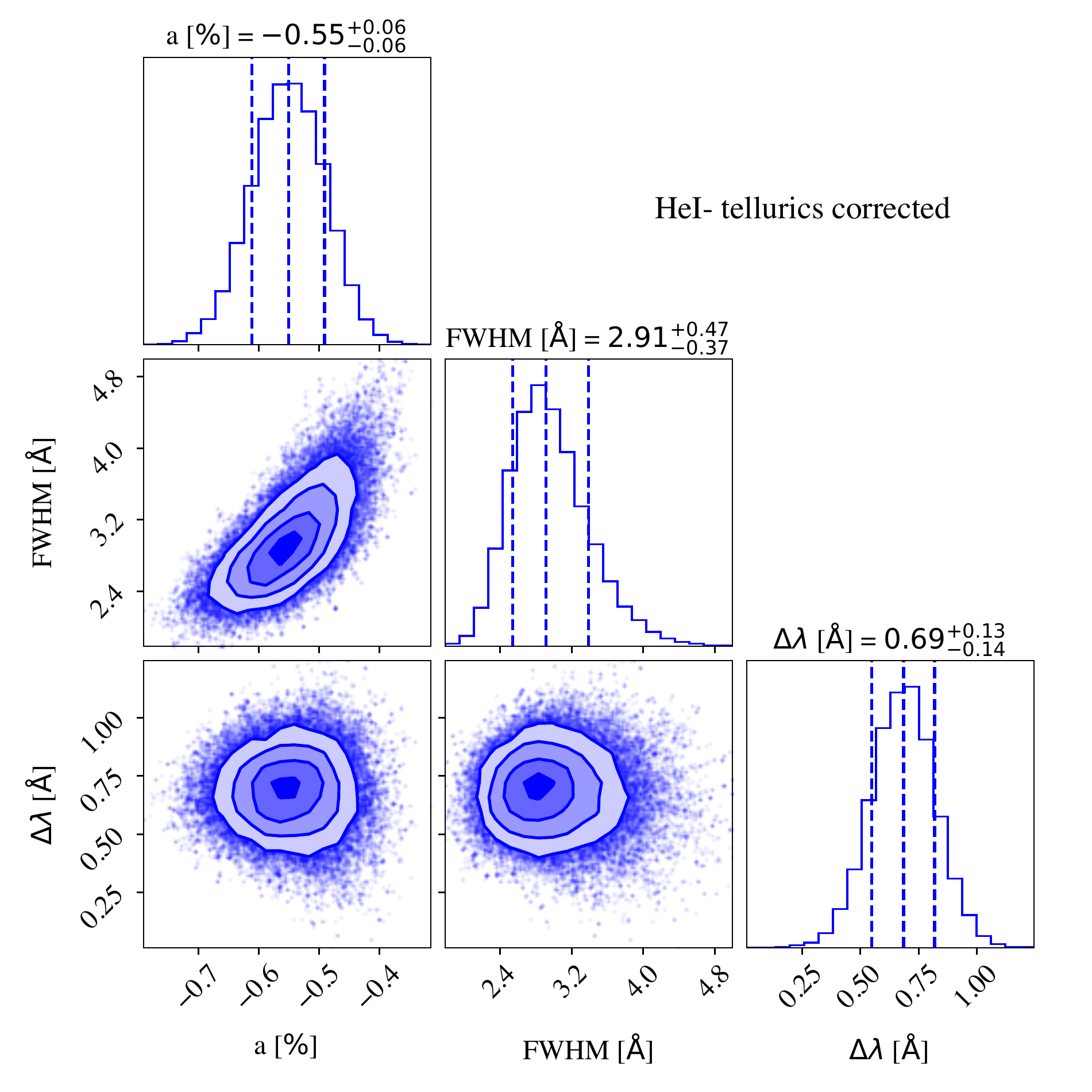}
\includegraphics[width=0.45\textwidth]{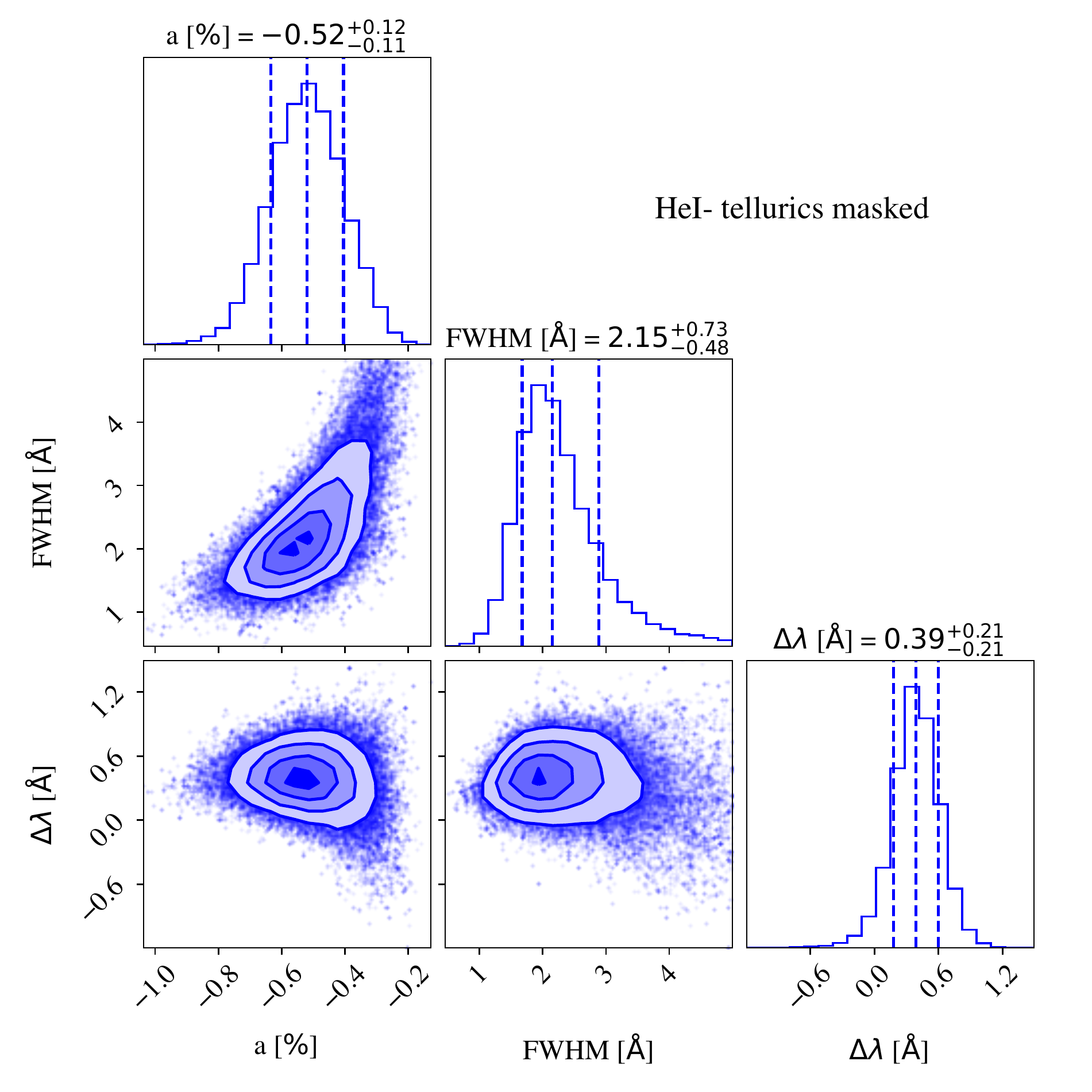}

\caption{MCMC probability distribution of the Gaussian profile parameters for the \ion{Ca}{ii} IRT $\lambda$8500\,{\AA} and $\lambda$8664\,{\AA} (\textit{top row}), and the \ion{He}{i} $\lambda$10830\,{\AA} (corrected and masked tellurics, \textit{bottom row}) absorption lines observed in the transmission spectrum of WASP-76b. The distributions were obtained using $10^4$ steps and $10$ walkers.}
\label{fig:corner}
\end{figure*}


\clearpage

\subsection{EMC distributions} \label{app:emc}

\begin{figure*}[h]
\centering
\includegraphics[width=0.31\textwidth]{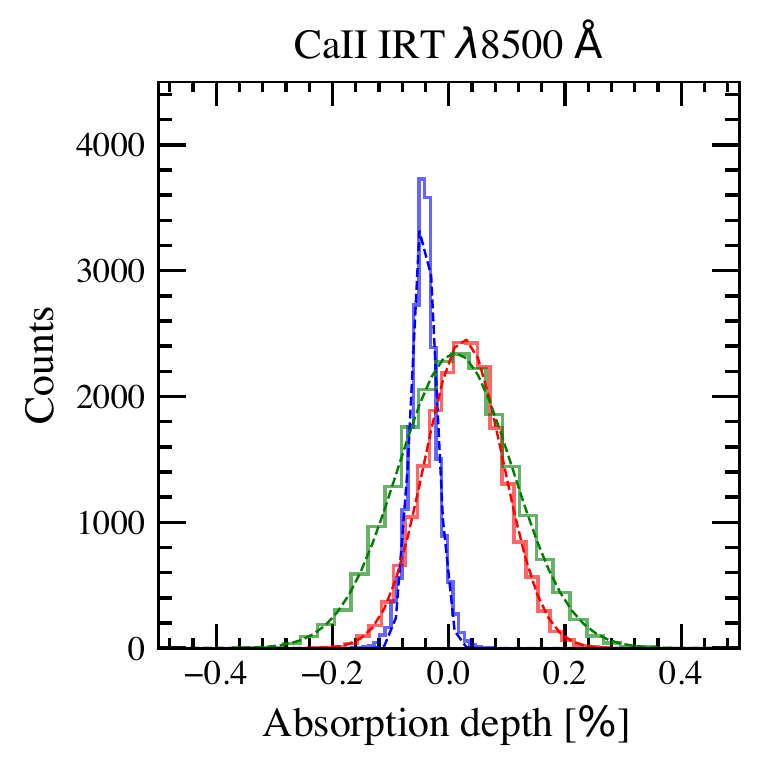}
\includegraphics[width=0.31\textwidth]{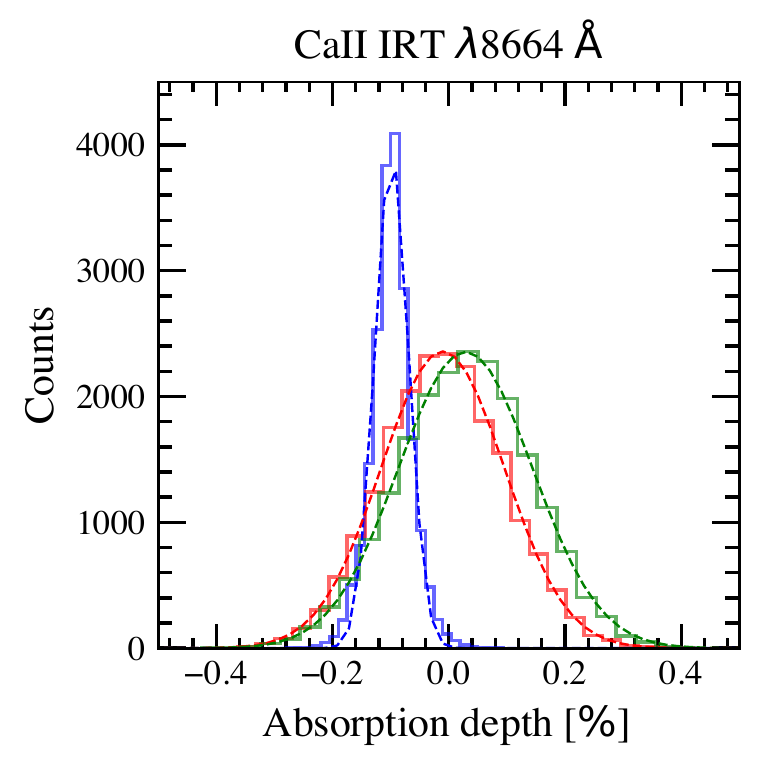}
\includegraphics[width=0.31\textwidth]{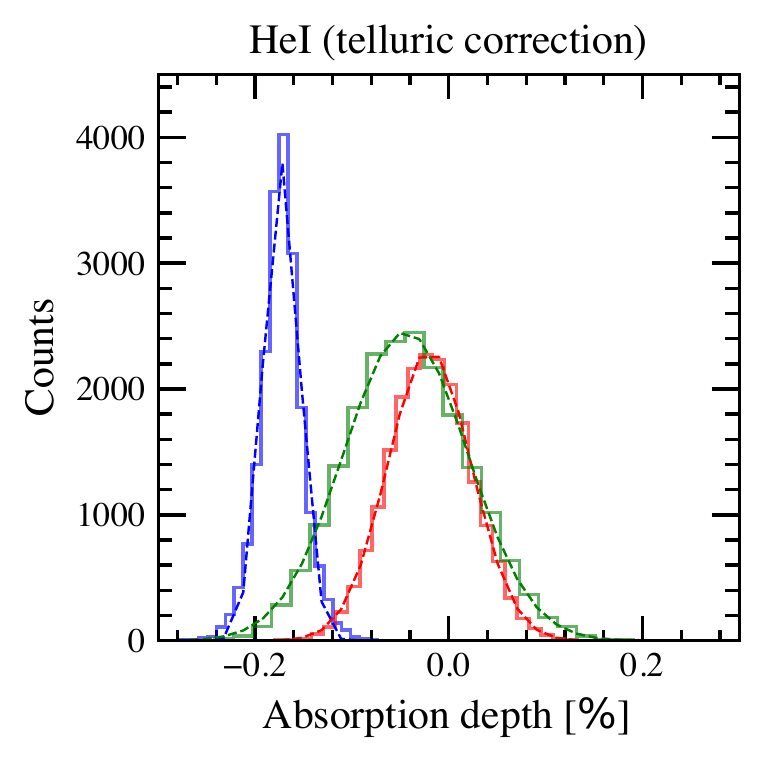}
\caption{EMC distributions around the \ion{Ca}{ii} IRT lines at 8500\,{\AA} and 8664\,{\AA}, and the \ion{He}{i} lines, for both nights combined. The distributions are obtained using $20~000$ iterations and a $2$\,{\AA} bandwidth for the \ion{Ca}{ii} IRT lines and $5$\,{\AA} bandwidth for the \ion{He}{i}, to cover the overall absorption feature observed in the transmission spectrum. The results of the `in-in' sample are shown in red, the `out-out' in green, and the `in-out' in blue. The dashed lines show the best-fit Gaussian profiles of the distributions.}
\label{fig:emc1}
\end{figure*}

\end{appendix}

\end{document}